\documentclass[12pt,reqno]{amsart}

\usepackage{amsmath,amsfonts,amsthm,amssymb}

\usepackage[letterpaper,height=20cm,top=24mm,bottom=24mm,left=26mm,right=26mm]{geometry}

\usepackage[numbers,sort]{natbib}
\numberwithin{equation}{section}
\newcommand{\I}{\mathrm{i}}
\newcommand{\E}{\mathrm{e}}

\DeclareMathOperator{\sign}{sgn}
\DeclareMathOperator{\Tr}{Tr}

\DeclareMathOperator{\ch}{ch}
\DeclareMathOperator{\erfc}{erfc}

\DeclareMathDelimiter{\Norm}{\mathord}{largesymbols}{"3E}{largesymbols}{"3E}

\DeclareMathOperator{\Wron}{Wr}
\begin{document}
\baselineskip 16pt
\parskip 8pt
\sloppy

%%%%%%%%%%%%%%%%%% TITLE %%%%%%%%%%%%%%

\title{
  Superconformal Algebras and Mock Theta Functions
}

%%%%%%%%%%%%%%%%%%%%%%% AUTHOR(S) %%%%%%%%%%%%%%%%%%%

\author[T. Eguchi]{Tohru \textsc{Eguchi}}
\author[K. Hikami]{Kazuhiro \textsc{Hikami}}

%%%%%%%%%%%%%%%%%%%%%%% ADDRESS %%%%%%%%%%%%%%%%%%%%

\address{Yukawa Institute for Theoretical Physics, Kyoto University,
  Kyoto 606--8502, Japan}
\email{
  \texttt{eguchi@yukawa.kyoto-u.ac.jp}
}

\address{Department of Mathematics, 
  Naruto University of Education,
  Tokushima 772-8502, Japan.}

\email{
  \texttt{hikami@naruto-u.ac.jp}
%  \texttt{KHikami@gmail.com}
}

%   \address{Department of Physics, Graduate School of Science,
% %    \\
%     University of Tokyo,
% %    \\
%     Hongo 7--3--1, Bunkyo, Tokyo 113--0033,   Japan.
%     }
% %     \\
    
% %    \urladdr{http://gogh.phys.s.u-tokyo.ac.jp/{\textasciitilde}hikami/}
%     \urladdr{\url{http://gogh.phys.s.u-tokyo.ac.jp/~hikami/}}

%     \email{\texttt{hikami@phys.s.u-tokyo.ac.jp}}

%%%%%%%%%%%%%%%%%%%%%% DATE %%%%%%%%%%%%%%%%%%%%%%%%%%%
%(Received: \hspace{40mm})

%\vspace{18pt}
%\date{\today}
\date{December 5, 2008. Revised on February 27, 2009}

%%%%%%%%%%%%%%%%%%%%%% ABSTRACT %%%%%%%%%%%%%%%%%%%%%%
\begin{abstract}

It is known that  characters of BPS representations of  extended
superconformal algebras do not have good modular properties due to
extra singular vectors coming from the BPS condition.  In order to
improve their modular properties we apply  the method of Zwegers which
has recently been developed to analyze modular properties of mock
theta functions. We consider the case of $\mathcal{N}=4$
superconformal algebra at general levels and obtain the decomposition
of characters of BPS representations into a sum of simple Jacobi forms
and an infinite series of non-BPS representations.

We apply our method to study elliptic genera of hyper-K\"ahler
manifolds in higher dimensions. In particular we determine the
elliptic genera  in the case of complex 4 dimensions of the Hilbert
scheme of points on K3 surfaces  $K^{[2]}$ and complex tori
$A^{[[3]]}$.
\end{abstract}

%%%%%%%%%%%%%%%%%%%%%%% Key Words %%%%%%%%%%%%%%%%%%%%%%%%%%


\subjclass[2000]{
}

%%%%%%%%%%%%%%%%%%%%%%%%%%%%%%%%%%%%%%%%%%%%%%%%%%%%%%%%%
%\newpage

%%\renewcommand{\thefootnote}{\arabic{footnote}}
\maketitle
%%%%%%%%%%%%%%%%%%%%%%%%%%%%%%%%%%%%

\section{Introduction}
Superconformal field theory (SCFT) in two-dimensions
provides a fundamental tool
in describing string theory compactified on some target
manifold $X$. One considers SCFT with 
an extended world-sheet $\mathcal{N}=2$ or $4$ supersymmetry
depending on whether the manifold $X$ is K\"ahler or
hyper-K\"ahler. It is well-known that the special feature
of SCFT's with extended SUSY 
is the existence of short or BPS representations 
in addition to the long or non-BPS representations. BPS
representations appear at special values of conformal
dimensions $h$ and describe massless states in 
compactified string theory which encode the geometrical
information of  target manifolds.  On the other hand,
non-BPS representations appear at continuous values of $h$
and correspond to massive excitations in string theory. As
we shall see, characters of BPS representations in general
do not have a simple transformation law under modular
transformations: this is due to the   
denominator factors which come from the BPS condition that
the supercharges annihilate BPS states.

In this paper we would like to propose a method which
replace   
characters of BPS representations by some Jacobi forms,
i.e. forms with good modular properties obtained by adding
an infinite series of characters of non-BPS
representations. This construction is based on an analogy
with Zwegers' treatment of mock theta functions where mock
theta functions are replaced by (real analytic) Jacobi
forms by the addition of Eichler integral of
suitable ``shadows'' of
mock theta functions.   

In the following we restrict our discussions to the case
of
$\mathcal{N}=4$ SCFT, however, very similar structures emerge also
in $ \mathcal{N}=2$ case.

Study of the $\mathcal{N}=4$ superconformal
field theory has been initiated some time ago
in~\cite{EgucTaor86a,EgucTaor88a,EgucTaor88b}.
$\mathcal{N}=4$ superconformal algebra (SCA) contains, besides the
energy-momentum tensor,
4 supercurrents and a triplet of currents which constitute the affine Lie algebra 
$SU(2)_k$.
There exits two types of representations, i.e. BPS and
non-BPS representation. We  
call their character formulas as massless and massive
characters, respectively.
Complexity of modular properties of massless characters
has been noticed in~\cite{EguOogTaoYan89a,EgucSugaTaor07a}
and 
the level-$k=1$ case has been considered in
detail in connection
with the geometry of
$K3$ surface.

The difficulty of the massless
character comes from the fact that it has the form of the
Lerch sum as in (\ref{define_massless_ch}).
Lerch sum has been long studied in analytic number theory
in connection with the so-called mock theta functions 
which first appeared in Ramanujan's last letter to Hardy in 1920
(see \emph{e.g.}, Ref.~\cite{GEAndre89a}).
Although mock theta function is not modular, it has a
quasi modular property as was first shown by
Watson~\cite{GWats36}.
Underlying intrinsic mathematical structure of these 
functions has recently been clarified by
Zwegers in his thesis~\cite{Zweg02Thesis};
the mock theta function is a holomorphic part of the
harmonic
Maass form with weight $1/2$, which is a Jacobi form and 
an eigenfunction of the second order differential
operator.
As an application, Bringmann and Ono resolved a
rank-generating problem of
integer partitions~\cite{FJDyson44} by constructing the
Poincar{\'e}--Maass
series~\cite{BrinKOno06a}.

The purpose of this paper is to apply Zwegers' method to 
the $\mathcal{N}=4$ superconformal algebras at general level-$k$ and
construct a Jacobi form for 
each massless representation. 
We show that the construction of Jacobi form is useful in
decomposing elliptic
genus in terms of the irreducible superconformal
representations. As an application we will determine the elliptic genera of 
hyper-K\"ahler manifolds in complex 4-dimensions, in particular the Hilbert scheme of points 
on $K3$ surfaces $K3^{[2]}$ and complex tori $A^{[[3]]}$. 

This paper is organized as follows.
In Section~\ref{sec:SCFT},
we briefly review the representation theory and character
formulas of $ \mathcal{N}=4$ SCA.
We reconsider  the level-1 case 
using Zwegers' formalism  
in Section~\ref{sec:level-1}.
In Section~\ref{sec:higher-level} we  study in detail the
higher-level $ \mathcal{N}=4$ SCFT and present elliptic genera
of complex 4-dimensional 
hyper-K\"ahler manifolds. 
Definitions of theta functions and numerical data of some
expansion coefficients are relegated to Appendices A and
B.

%%%%%%%%%%%%%%%%%%%%%%%%
\section{Superconformal Algebra and Elliptic Genus}
\label{sec:SCFT}
\subsection{Character of Superconformal Algebra}
The $\mathcal{N}=4$ superconformal algebra
%\cite{AdeBri76a}
with central charge
$c= 6\, k$
contains a level-$k$ affine SU(2) algebra.
The highest weight states are labeled by the conformal dimension
$h$ and isospin $\ell$, satisfying
$0 \leq \ell \leq k/2$
with $\ell \in \mathbb{Z}/2$.
Due to the unitarity, we have a bound on conformal dimension in R and NS sector as
\begin{equation*}
  \begin{cases}
    \text{R} : &
    \displaystyle
    h \geq \frac{k}{4}  ,
    \\[2mm]
    \text{NS}    : &
    h \geq \ell .
  \end{cases}
\end{equation*}
In each of these sectors,
there exist two types of 
representations~\cite{EgucTaor86a,EgucTaor88a,EgucTaor88b};
\begin{itemize}

\item massive (non-BPS) representation,
  \begin{equation*}
    \begin{cases}
      \text{R}: &
      h > \frac{k}{4},
      \qquad
      \ell=\frac{1}{2},1,  \dots, \frac{k}{2}, 
      \\
      \text{NS} : &
      h > \ell,
      \qquad
      \ell=0,\frac{1}{2}, \dots, \frac{k-1}{2}, 
      \qquad
    \end{cases}
  \end{equation*}

\item massless (BPS) representation,
  \begin{equation*}
    \begin{cases}
      \text{R} : &
      h = \frac{k}{4},
      \qquad
      \ell=0,\frac{1}{2}, \dots, \frac{k}{2}, 
      \\
      \text{NS} : &
      h = \ell,
      \qquad
      \ell=0,\frac{1}{2}, \dots, \frac{k}{2} .
    \end{cases}
  \end{equation*}

\end{itemize}
The character is a  trace over the representation space 
\begin{equation}
  \ch_{k,h,\ell}(z;\tau)
  =
  \Tr_{\mathcal{H}}
  \left(
    \E^{2 \pi \I z T_0^3} \,
    q^{L_0 - \frac{c}{24}}
  \right)  ,
\end{equation}
where $q=\E^{2 \pi \I \tau}$ with
$\tau \in \mathbb{H}$ as usual. $L_0$ and $T^3_0$ denote the zero mode of Virasoro operator and the 3rd component of the $SU(2)$ current, respectively.
The characters in the Ramond sector, for instance, are computed
as~\cite{EgucTaor86a,EgucTaor88a}
\begin{itemize}
\item massive character,
  \begin{equation}
    \label{massive_character}
    \ch^R_{k,h,\ell}(z;\tau)
    =
    q^{h - \frac{\ell^2}{k+1} - \frac{k}{4}} \,
    \frac{
      \left[ \theta_{10}(z;\tau) \right]^2
    }{
      \left[ \eta(\tau) \right]^3
    } \,
    \chi_{k-1, \ell - \frac{1}{2}}(z;\tau) ,
  \end{equation}

\item massless character,
  \begin{multline}
    \ch^R_{k,\frac{k}{4},\ell}(z;\tau)
    =
    \frac{\I}{\theta_{11}(2 z; \tau)} \cdot
    \frac{
      \left[ \theta_{10}(z;\tau) \right]^2
    }{
      \left[ \eta(\tau) \right]^3
    }
    \\
    \times
    \sum_{\varepsilon=\pm 1} \sum_{m \in \mathbb{Z}}
    \varepsilon \,
    \frac{
      \E^{4 \pi \I \varepsilon \left( (k+1)m+\ell \right) z}
    }{
      \left(
        1+ \E^{- 2 \pi \I \varepsilon z} \, q^{-m}
      \right)^2
    } \,
    q^{(k+1) m^2 + 2 \ell m} .
    \label{define_massless_ch}
  \end{multline}
\end{itemize}
See Appendix~\ref{sec:theta} for definitions of theta
functions. $\chi_{k,\ell}(z;\tau)$ denotes the character of the spin
$\ell$ representation of $SU(2)_k$ algebra. 
Note that the denominator  in the massless character 
originates from the BPS condition that the supercharge annihilates the BPS 
states.
If we ignore the denominator $(1+\E^{-2\pi \I\varepsilon z}\,q^{-m})^2$
in the massless character, it takes the same form as the massive
character at $h=k/4$. 
We can check directly that
the massless characters~\eqref{define_massless_ch}
satisfy a recursion relation
\begin{multline}
  \label{ch_recursion}
  \ch_{k,\frac{k}{4},\ell}^R(z;\tau)
  + 
  2\, \ch_{k,\frac{k}{4},\ell-\frac{1}{2}}^R(z;\tau)
  +
  \ch_{k,\frac{k}{4},\ell-1}^R(z;\tau)
  \\
  =
  q^{ - \frac{\ell^2}{k+1}} \,
  \frac{
    \left[ \theta_{10}(z;\tau) \right]^2
  }{
    \left[ \eta(\tau) \right]^3
  } \,
  \chi_{k-1, \ell - \frac{1}{2}}(z;\tau) .
\end{multline}
Note that the right hand side
corresponds to the  character of
massive representation in the limit,
\begin{equation*}
  \lim_{h \searrow \frac{k}{4}}
  \ch_{k,h,\ell}^R(z;\tau)
  =
  q^{ - \frac{\ell^2}{k+1}} \,
  \frac{
    \left[ \theta_{10}(z;\tau) \right]^2
  }{
    \left[ \eta(\tau) \right]^3
  } \,
  \chi_{k-1, \ell - \frac{1}{2}}(z;\tau) ,
\end{equation*}
and \eqref{ch_recursion} shows how the non-BPS representation
decomposes into a sum of BPS representations at the unitarity bound.

The characters in the NS sector are given from those in the R sector by
the spectral flow as
\begin{equation}
  \ch_{k,h+\frac{\ell}{2}+\frac{k}{4},\frac{k}{2}-\ell}^{NS}(z;\tau)
  =
  q^{\frac{k}{4}} \, \E^{2 \pi \I k z} \,
  \ch_{k,h,\ell}^R
  \left( z+\frac{\tau}{2};\tau \right) .
\end{equation}
Characters of other sectors are also defined by
\begin{align}
  \ch_{k,h,\ell}^{\widetilde{R}}(z;\tau)
  & =
  \ch_{k,h,\ell}^{R}
  \left( z+\frac{1}{2} ; \tau \right) ,
  \\[2mm]
  \ch_{k,h,\ell}^{\widetilde{NS}}(z;\tau)
  & =
  \ch_{k,h,\ell}^{NS}
  \left( z+\frac{1}{2} ; \tau \right) .
\end{align}

%%%%%%%%%%%
\subsection{Elliptic Genus}
When the superconformal field theory  is interpreted as a nonlinear sigma
model on the Calabi--Yau manifold $X$ with
a  complex dimension  $c/3$,
the elliptic genus of $X$ is identified with~\cite{EWitt87a}
\begin{equation}
  Z_X(z;\tau)
  =
  \Tr_{\mathcal{H}^R \otimes \mathcal{H}^R}
  (-1)^F \, \E^{2 \pi \I T_0^3 z} \,
  q^{L_0 - \frac{c}{24}} \,
  \overline{q}^{\overline{L}_0 - \frac{c}{24}} ,\label{elliptic-genus}
\end{equation}
where
$(-1)^F=\E^{\pi \I (T_0^3 - \overline{T}_0^3)}$,
and $\mathcal{H}^R$ denotes Hilbert space of the Ramond sector.
Due to the supersymmetry, the elliptic genus is independent of
$\overline{q}$,
and only the ground states contributes in the right-moving sector.
The fundamental structure of the elliptic genus
$Z_X(z;\tau)$ is the modular
property~\cite{EichZagi85,KawaYamaYang94a};
\begin{equation}
  \begin{gathered}
    Z_X
    \left(
      \frac{z}{\tau} ; - \frac{1}{\tau}
    \right)
    =
    \E^{2 \pi \I \frac{c}{6} \frac{z^2}{\tau}} \,
    Z_X(z;\tau) ,
    \\[2mm]
    Z_X(z;\tau+1) =
    Z_X(z;\tau) ,
    \\[2mm]
    Z_X(z+1;\tau)
    =
    (-1)^{\frac{c}{3}} \, Z_X(z;\tau),
    \\[2mm]
    Z_X(z+\tau ; \tau)
    =
    (-1)^{\frac{c}{3}} \,
    q^{-\frac{c}{6}}    \, \E^{-2 \pi \I \frac{c}{3} z} \,
    Z_X(z;\tau) .
  \end{gathered}
\end{equation}
It is known that the elliptic genus
at special values of $z$ gives classical topological invariants of
$X$~\cite{EguOogTaoYan89a};
\begin{equation}
  \begin{aligned}
    Z_X(z=0;\tau)
    & =
    \chi_X, 
    \\[2mm]
    Z_X \left( z=\frac{1}{2};\tau\right)
    & =
    \sigma_X + \mathcal{O}(q) ,
    \\[2mm]
    q^{\frac{c}{12}} \,
    Z_X \left( z=\frac{1+\tau}{2} ; \tau \right)
    & =
    \widehat{A}_X + \mathcal{O}(q),
  \end{aligned}
\end{equation}
where $\chi_X$, $\sigma_X$, and $\widehat{A}_X$ are respectively
the Euler characteristic,
the Hirzebruch signature, and the $\widehat{A}$-genus of $X$.
%%%%%%%%%%%%%%%%
\section{%\mathversion{bold}
  Level $k=1$ Superconformal Algebra Revisited}
\label{sec:level-1}

The massless character in the $\widetilde{R}$-sector is read
from~\eqref{define_massless_ch} as
\begin{gather}
  \label{massless_ch_1}
  \ch_{k=1,h=\frac{1}{4},\ell=0}^{\widetilde{R}}(z;\tau)
  =
  \frac{\I}{\theta_{11}(2 z;\tau)} \cdot
  \frac{
    \left[ \theta_{11}(z;\tau) \right]^2
  }{
    \left[ \eta(\tau) \right]^3
  }
  \sum_{m \in \mathbb{Z}}
  q^{2 m^2} \, \E^{8 \pi \I m z} \,
  \frac{1 + \E^{2 \pi \I z} \, q^m}{
    1 - \E^{2 \pi \I z} \, q^m} ,
\end{gather}
and the identity~\eqref{ch_recursion} reduces to
\begin{equation}
  \label{recursion_ch_R_prime}
  \ch^{\widetilde{R}}_{k=1,h=\frac{1}{4},\ell=\frac{1}{2}}(z;\tau)
  +
  2 \,
  \ch^{\widetilde{R}}_{k=1,h=\frac{1}{4},\ell=0}(z;\tau)
  =
  q^{-\frac{1}{8}} \,
  \frac{
    \left[ \theta_{11}(z;\tau) \right]^2
  }{
    \left[ \eta(\tau) \right]^3
  } .
\end{equation}
The character~\eqref{massless_ch_1}
is   known to  be rewritten as~\cite{EgucTaor88b}
\begin{equation}
  \label{massless_ch_M}
  \ch_{k=1,h=\frac{1}{4},\ell=0}^{\widetilde{R}}(z;\tau)
  =
  \frac{
    \left[    \theta_{11}(z;\tau) \right]^2
  }{
    \left[ \eta(\tau) \right]^3
  } \,
  \mu(z; \tau) ,
\end{equation}
where
\begin{equation}
  \label{define_M}
  \mu(z;\tau)
  =
  \frac{
    \I \, \E^{\pi \I z} 
  }{
    \theta_{11}(z;\tau)}
  \sum_{n \in \mathbb{Z}}
  (-1)^n \,
  \frac{
    q^{\frac{1}{2} n(n+1)} \, \E^{2 \pi \I  n z}
  }{
    1 - q^n \, \E^{2 \pi \I z}
  } .
\end{equation}
We reconsider the  modular properties of $\mu(z;\tau)$  using Zwegers'
approach.

\subsection{Harmonic Maass Form}

Zwegers defines the function~\cite{Zweg02Thesis}
\begin{equation}
  \mu(u,v;\tau)
  =
  \frac{\I \, \E^{\pi \I u}}{
    \theta_{11}(v ; \tau)
  }
  \sum_{n \in \mathbb{Z}}
  \frac{
    \left(- \E^{2 \pi \I v} \right)^n \, q^{\frac{1}{2}n(n+1)}
  }{
    1-\E^{2 \pi \I u} \, q^n
  } ,
  \label{define_mu}
\end{equation}
which is known as the Lerch sum
(or, generalized Lambert series). $\mu(z;\tau)$ of (\ref{define_M}) and 
$\mu(u,v;\tau)$ 
are simply related as 
$\mu(u=z,v=z;\tau)=\mu(z;\tau)$. 
Zwegers showed that,
under the $S$-transformation, $\mu(u,v;\tau)$  behaves as
\begin{equation}
  \mu(u,v; \tau)
  + 
  \sqrt{\frac{\I}{\tau}} \,
  \E^{\pi \I \frac{(u-v)^2}{\tau}} \,
  \mu
  \left(
    \frac{u}{\tau} , \frac{v}{\tau} ; - \frac{1}{\tau}
  \right)
  =
  \frac{1}{2} \, M(u-v; \tau) ,
  \label{S_mu}
\end{equation}
where $M(x;\tau)$ is the Mordell integral~\cite{LJMorde33a} defined by
\begin{equation}
  M(x;\tau)
  =
  \int_{-\infty}^\infty
  \frac{
    \E^{\pi \I \tau z^2 - 2 \pi x z}
  }{\cosh(\pi \, z)} \,
  \mathrm{d} z .
\end{equation}
He  further introduced a
non-holomorphic partner of $\mu(u,v; \tau)$ by
\begin{multline}
  R(z;\tau)
  =
  \sum_{n \in \mathbb{Z}} (-1)^n \,
  \left[
    \sign\left(n+\frac{1}{2}\right)
    -
    E\left(
     \left( n + \frac{1}{2} + \frac{\Im z}{\Im \tau} \right) \,
     \sqrt{2 \, \Im \tau}
    \right)
  \right]
  \\
  \times
  \E^{-2 \pi \I \left( n + \frac{1}{2} \right) z} 
  \, q^{-\frac{1}{2} \left( n + \frac{1}{2} \right)^2} ,
\end{multline}
where $E(z)$ is the error function defined by
\begin{align}
  E(z)
  = 2 \int_0^z \E^{- \pi u^2} \mathrm{d} u 
  =
  1 - \erfc \left(  \sqrt{\pi} \, z \right) .
\end{align}
The non-holomorphic function
is shown to   fulfill the $S$-transformation 
\begin{equation}
  R(u;\tau)
  + \sqrt{\frac{\I}{\tau}} \,
  \E^{\pi \I \frac{u^2}{\tau}} \,
  R \left(
    \frac{u}{\tau} ; - \frac{1}{\tau}
  \right)
  =
  M(u;\tau) .
\end{equation}
Zwegers combined these two functions to define
\begin{equation}
  \widehat{\mu}(u,v;\tau)
  =
  \mu(u,v;\tau) -
  \frac{1}{2} R(u-v; \tau) ,
\end{equation}
as a completion of the Lerch sum $\mu(u,v;\tau)$.
The Mordell integral
$M(u;\tau)$
disappears in the $S$-transformation of
$\widehat{\mu}(u,v;\tau)$,
and
it behaves like a  2-variable Jacobi form with weight
$1/2$~\cite{EichZagi85}
\begin{equation}
  \begin{aligned}
    \widehat{\mu}(u,v;\tau)
    & =
    - \sqrt{\frac{\I}{\tau}}  \,  \E^{\pi \I \frac{(u-v)^2}{\tau}} \,
    \widehat{\mu}
    \left(
      \frac{u}{\tau}, \frac{v}{\tau}; - \frac{1}{\tau}
    \right)
    \\
    & =
    \E^{\pi \I/4} \, \widehat{\mu}(u,v; \tau+1) .
  \end{aligned}
\end{equation}

In view of~\eqref{define_M}
and~\eqref{define_mu},
we see that the  massless character~\eqref{massless_ch_M}
is proportional to the holomorphic part of the
real analytic modular form
\begin{equation}
  \label{define_1_M_hat}
  \widehat{\mu}(z;\tau)
  =
  \mu(z;\tau)
  - \frac{1}{2} \, R(0;\tau) ,
\end{equation}
whose transformation formulae are
\begin{equation}
  \label{modular_M_hat}
  \begin{gathered}
    \widehat{\mu}(z;\tau)
    =
    -
    \sqrt{\frac{\I}{\tau}} \,
    \widehat{\mu}\left(
      \frac{z}{\tau} ; - \frac{1}{\tau}
    \right) ,
    \\[2mm]
    \widehat{\mu}(z;\tau+1) =
    \E^{-\frac{1}{4} \pi \I} \,
    \widehat{\mu}(z;\tau) ,
    \\[2mm]
    \widehat{\mu}(z+1;\tau)
    =
    \widehat{\mu}(z+\tau;\tau)
    =
    \widehat{\mu}(z;\tau)  .   
  \end{gathered}
\end{equation}
It is possible to show
that the non-holomorphic part
$R(0;\tau)$ in~\eqref{define_1_M_hat}
is an incomplete period integral of the
modular form
$\left[ \eta(\tau) \right]^3$
with weight $3/2$;
\begin{equation}
  \label{R0_integral}
  \I \, R(0;\tau)
  =
  \int_{- \overline{\tau}}^{\I \infty}
  \frac{
    \left[ \eta(z) \right]^3
  }{
    \sqrt{
      \frac{z+\tau}{\I}
    }} \,
  \mathrm{d} z ,
\end{equation}
which shows that
the massless
character~\eqref{massless_ch_M}
of the level-$1$
superconformal algebra has 
a modular form
$\left[ \eta(\tau) \right]^3$  as a ``shadow''
in the sense of Zagier~\cite{Zagier08a}.
We then see that the real analytic function $\widehat{\mu}(z;\tau)$
satisfies
\begin{equation}
  \frac{\partial}{\partial \overline{\tau}}
  \widehat{\mu}(z;\tau)
  =
  \frac{
    \I}{2} \,
  \frac{
    \left[ \eta( - \overline{\tau}) \right]^3
  }{
    \sqrt{2 \, \Im \tau}
  } .
\end{equation}
To conclude, we see that the function $\widehat{\mu}(z;\tau)$ is a
harmonic Maass form being a solution of the differential equation
\begin{equation}
  \label{differential_Maass}
  \left( \Im \tau \right)^{\frac{3}{2}}
  \frac{\partial}{\partial \tau}  \sqrt{\Im \tau} 
  \frac{\partial}{\partial \overline{\tau}} \,
  \widehat{\mu}(z;\tau)
  = 0 ,
\end{equation}
which,
with $\tau=u+\I \, v$, reduces to
\begin{equation}
  \left[
    -v^2 \,
    \left(
      \frac{\partial^2}{\partial u^2}
      + \frac{\partial^2}{\partial v^2}
    \right)
    + 
    \frac{\I}{2} \, v \,
    \left(
      \frac{\partial}{\partial u}
      + \I \,
      \frac{\partial}{\partial v}
    \right)
  \right] \,
  \widehat{\mu}(z;\tau)
  = 0 .
\end{equation}

We notice that
the  formula under the $S$-transformation~\eqref{S_mu}
gives~\cite{EgucTaor88b}
\begin{equation}
  \label{k1_character_S}
  {\mu\over \eta}(z; \tau)
  +
  {\mu \over \eta}\left(
    \frac{z}{\tau}; - \frac{1}{\tau}
  \right) 
  =
  \frac{1}{\eta(\tau)} \,
  \int_{-\infty}^\infty
  \frac{\E^{\pi \I \tau x^2}}{2 \cosh(\pi \, x)} \, \mathrm{d} x .
\end{equation}
%where
%we  normalize the level-1 massless character to
%define
%\begin{align}
%  \label{define_1_H}
%  H(z; \tau)
 % & =
 % \frac{M(z;\tau)}{
 %   \eta(\tau)
 % } .
%\end{align}

% Corresponding to~\eqref{define_1_M_hat}
% the function $H(z;\tau)$ is the holomorphic part of  the real-analytic
% modular form defined by
% \begin{equation}
%   \label{define_H_hat}
%   \widehat{H}(z;\tau) = \frac{\widehat{M}(z;\tau)}{\eta(\tau)} .
% \end{equation}
% whose modular  transformation formulae are given by
% \begin{equation}
%   \label{modular_H_hat}
%   \begin{gathered}
%     \widehat{H}(z;\tau)
%     = -
%     \widehat{H}
%     \left( \frac{z}{\tau} ; - \frac{1}{\tau}  \right) ,
%     \\[2mm]
%  %
%     \widehat{H}(z;\tau+1) = \E^{- \frac{\pi \I}{3}} \,
%     \widehat{H}(z;\tau) ,
%     \\[2mm]
%     % 
%     \widehat{H}(z+1;\tau)
%     = \widehat{H}(z+\tau; \tau)
%     = \widehat{H}(z;\tau) .
%   \end{gathered}
% \end{equation}

%%%%
\subsection{Character Decomposition of Elliptic Genus}

Let the function $J(z;w;\tau)$  be
\begin{align}
  \label{define_lowest_T}
  J(z; w; \tau)
  & =
   \frac{
     \left[ \theta_{11}(z;\tau) \right]^2
   }{
     \left[ \eta(\tau) \right]^3
   }
   \,
  \left(
    \widehat{\mu}(z;\tau) -
    \widehat{\mu}(w;\tau) 
  \right)
  \\
  & =
  \frac{
    \left[ \theta_{11}(z;\tau) \right]^2
  }{
    \left[ \eta(\tau) \right]^3
  }
  \,
  \left(
    {\mu}(z;\tau) -
    {\mu}(w;\tau) 
  \right)
  \nonumber
  \\
  & = 
  \ch_{k=1,h=\frac{1}{4},\ell=0}^{\widetilde{R}}(z;\tau)
  -
  \frac{
    \left[ \theta_{11}(z;\tau) \right]^2
  }{
    \left[ \eta(\tau) \right]^3
  }  \,
  {\mu}(w;\tau)   ,
  \nonumber
\end{align}
where the second equality follows from
the $z$-independence of $R(0;\tau)$~\eqref{define_1_M_hat}.
We see that
\begin{equation}
  J(w;w;\tau) = 0 ,
\end{equation}
and that
$J(0;w;\tau)=1$.
By use of~\eqref{modular_M_hat},
we obtain the transformation formulae
%are like 2-variable Jacobi form
as follows;
\begin{equation}
  \begin{gathered}
    J(z;w;\tau)
    =
    \E^{-2 \pi \I \frac{z^2}{\tau}} \,
    J \left(
      \frac{z}{\tau} ; \frac{w}{\tau}; - \frac{1}{\tau}
    \right) ,
    \\[2mm]
    J(z+1; w; \tau)
    =
    J(z; w; \tau+1)
    = J(z; w+\tau; \tau)
    = J(z; w; \tau) ,
    \\[2mm]
    J(z+\tau; w; \tau)
    =
    q^{-1} \, \E^{- 4 \pi \I z} \,
    J(z;w;\tau) .
  \end{gathered}
\end{equation}
The second equality of~\eqref{define_lowest_T} indicates that the
function $J(z;w;\tau)$ is a holomorphic function;
when we fix $w$ to be specific values $w=\frac{1}{2}$,
$\frac{1+\tau}{2}$, and $\frac{\tau}{2}$,
those modular properties show
\begin{equation}
  \label{1_T_theta_2}
    J \left( z;\frac{1}{2}; \tau \right)
    =
    \left(
      \frac{\theta_{10}(z;\tau)}{
        \theta_{10}(0;\tau)}
    \right)^2\hskip-3mm  ,
    J \left( z;\frac{1+\tau}{2}; \tau \right)
     =
    \left(
      \frac{\theta_{00}(z;\tau)}{
        \theta_{00}(0;\tau)}
    \right)^2\hskip-3mm ,
    J \left( z;\frac{\tau}{2}; \tau \right)
   =
    \left(
      \frac{\theta_{01}(z;\tau)}{
        \theta_{01}(0;\tau)}
    \right)^2 \hskip-3mm .
\end{equation}
Lerch sums introduced in~\cite{EgucTaor88b} are given by the values of $\mu(w;\tau)$ at 
$w=1/2,1+\tau/2,\tau/2$
\begin{equation}
  \label{define_h_from_H}
  \begin{aligned}
    h_2(\tau)
    \equiv
    {\mu \over \eta} \left( \frac{1}{2};\tau \right)
    & =
    \frac{1}{
      \eta(\tau) \, \theta_{10}(0;\tau)}
    \sum_{n \in \mathbb{Z}}
    \frac{q^{\frac{1}{2} n(n+1)}}{
      1+ q^n} ,
    \\
    h_3(\tau)
    \equiv
    {\mu \over \eta} \left( \frac{1+\tau}{2}; \tau \right)
    & =
    \frac{1}{
      \eta(\tau) \, \theta_{00}(0;\tau)}
    \sum_{n \in \mathbb{Z}}
    \frac{q^{\frac{1}{2} n^2 - \frac{1}{8}}}{
      1+ q^{n-\frac{1}{2}}} ,
    \\
    h_4(\tau)
    \equiv
    {\mu \over \eta}\left(\frac{\tau}{2}; \tau\right)
    & =
    \frac{1}{
      \eta(\tau) \, \theta_{01}(0;\tau)}
    \sum_{n \in \mathbb{Z}} (-1)^n
    \frac{
      q^{\frac{1}{2}n^2 - \frac{1}{8}}}{
      1+ q^{n-\frac{1}{2}}} .
  \end{aligned}
\end{equation}
We then obtain the decomposition formula for the massless character
\begin{eqnarray}
 \ch_{k=1,h=\frac{1}{4},\ell=0}^{\widetilde{R}}(z;\tau)
&=& \left(
      \frac{\theta_{10}(z;\tau)}{
        \theta_{10}(0;\tau)}
    \right)^2 + h_2(\tau)   \frac{
    \left[ \theta_{11}(z;\tau) \right]^2
  }{
    \left[ \eta(\tau) \right]^2
  } , \nonumber \\
  &=& \left(
      \frac{\theta_{00}(z;\tau)}{
        \theta_{00}(0;\tau)}
    \right)^2 + h_3(\tau)   \frac{
    \left[ \theta_{11}(z;\tau) \right]^2
  }{
    \left[ \eta(\tau) \right]^2
  },  \label{decomposition-at-k=1} \\
  &= &\left(
      \frac{\theta_{01}(z;\tau)}{
        \theta_{01}(0;\tau)}
    \right)^2 + h_4(\tau)   \frac{
    \left[ \theta_{11}(z;\tau) \right]^2
  }{
    \left[ \eta(\tau) \right]^2
  }.  \nonumber 
    \end{eqnarray}

Identities~\eqref{decomposition-at-k=1} are used to express the elliptic
genus of K3 surface in terms of irreducible representations of $\mathcal {N}=4$ algebra.
The elliptic genus of K3 surface is known to be given
by~\cite{EguOogTaoYan89a,KawaYamaYang94a}
\begin{equation}
  \label{genus_K3}
  Z_{K3}(z;\tau)
  =
  8 \,
  \left[
    \left(
      \frac{\theta_{10}(z;\tau)}{\theta_{10}(0;\tau)}
    \right)^2
    +
    \left(
      \frac{\theta_{00}(z;\tau)}{\theta_{00}(0;\tau)}
    \right)^2
    +
    \left(
      \frac{\theta_{01}(z;\tau)}{\theta_{01}(0;\tau)}
    \right)^2
  \right] .
\end{equation}
Using~\eqref{decomposition-at-k=1}, we can rewrite
the elliptic
genus  as
\begin{equation*}
  Z_{K3}(z;\tau)
  =
  24 \,
  \ch_{k=1,h=\frac{1}{4},\ell=0}^{\widetilde{R}}(z;\tau)
  -
  8 \,
  \left[
    \frac{\theta_{11}(z;\tau)}{\eta(\tau)}
  \right]^2 \,
  \sum_{a=2,3,4} h_a(\tau) .
\end{equation*}
We note that
\begin{align*}
  8 \hskip-4mm \sum_{
    w \in \left\{
      \frac{1}{2} , \frac{1+\tau}{2} , \frac{\tau}{2}
    \right\}
  }\hskip-4mm 
  \mu(w; \tau)
  & =
  8 \, \eta(\tau) \hskip-2mm 
  \sum_{a=2,3,4}  \hskip-2mm h_a(\tau)
  \\
  &=
  q^{-\frac{1}{8}}\left(
    2 - \sum_{n \geq 1} A_n \, q^n
  \right) ,
\end{align*}
where $A_n$ are positive integers;
\begin{equation}
  \label{l1_coefficient_massive}
  \begin{array}{c|rrrrrrrrr}
    n & 1 & 2 & 3 & 4 & 5 & 6 & 7 & 8 & \cdots\\
    \hline
%     A_n& 45 & 231 & 770 & 2277 & 5796 & 13915 &
%    30843 & 65550 &\dots
    A_n& 90 & 462 & 1540 & 4554 & 11592 & 27830 &
    61686 & 131100 &\cdots
  \end{array}
\end{equation}
Using the relation~\eqref{recursion_ch_R_prime},
we obtain
\begin{equation}
  \label{decompose_K3}
  Z_{K3}(z;\tau)
  =
  20 \, \ch_{1,\frac{1}{4},0}^{\widetilde{R}}(z;\tau)
  -
  2 \,  \ch_{1,\frac{1}{4},\frac{1}{2}}^{\widetilde{R}}(z;\tau)
  +
  \sum_{n \geq 1} A_n \,
  \ch_{1,n+\frac{1}{4},\frac{1}{2}}^{\widetilde{R}}(z;\tau) ,
\end{equation}
where the first two terms are massless characters and the last term
gives an infinite sum of massive characters.
Note that the
right-moving sector in the elliptic genus (\ref{elliptic-genus}) gives
a degeneracy factor $2\ell+1$ for isospin $\ell$ representation. Hence
the net multiplicity of isospin $1/2$ massless representation in
(\ref{decompose_K3}) equals 1. Since isospin $1/2$ representation in R
sector flows to isospin $0$ in NS sector, (\ref{decompose_K3}) has the
multiplicity 1 for isospin $0$ or vacuum representation in NS sector.

%%%%%%%%%%%%%
\section{Higher Level Superconformal Algebra}
\label{sec:higher-level}

Massless  character at level-$k$ with  isospin-$0$
in the $\widetilde{R}$-sector is read from~\eqref{define_massless_ch}
as
\begin{equation}
  \label{level-k_ch_0}
  \ch_{k,\frac{k}{4},\ell=0}^{\widetilde{R}}(z;\tau)
  =
  \frac{\I}{\theta_{11}(2 \, z; \tau)} \cdot
  \frac{
    \left[ \theta_{11}(z;\tau) \right]^2
  }{
    \left[ \eta(\tau) \right]^3
  } \,
  \sum_{m \in \mathbb{Z}}
  \E^{4 \pi \I (k+1) m z} \, q^{(k+1) m^2} \,
  \frac{
    1+ \E^{2 \pi \I z} \, q^m}{
    1 - \E^{2 \pi \I z} \, q^m} ,
\end{equation}
and
the recursion relation~\eqref{ch_recursion} of
massless  characters becomes
\begin{multline}
  \label{recursion_massless_R_tilde}
  \ch^{\widetilde{R}}_{k,\frac{k}{4},\ell}(z;\tau)
  +
  2 \,
  \ch^{\widetilde{R}}_{k,\frac{k}{4},\ell-\frac{1}{2}}(z;\tau)
  +
  \ch^{\widetilde{R}}_{k,\frac{k}{4},\ell-1}(z;\tau)
  \\
  =
  (-1)^{2\ell+1} \,
  q^{- \frac{\ell^2}{k+1}} \,
  \frac{
    \left[ \theta_{11}(z;\tau) \right]^2
  }{
    \left[ \eta(\tau) \right]^3
  } \,
  \chi_{k-1,\ell - \frac{1}{2}}(z;\tau) .
\end{multline}
The  modular transformation properties of~\eqref{level-k_ch_0} are
studied \cite{EgucSugaTaor08a} by use of the formula of Miki
\cite{KMiki90a} and also the higher-level Appell
function~\cite{SemikTipunTaorm05a,Taorm08a}.
We shall  show that
these massless characters are
the mock theta functions whose ``shadow''~\cite{Zagier08a}
is $\Psi_{k+1}^{(a)}(\tau)$~\eqref{Def_Psi} related to the affine SU(2)
characters at level-$k$.
% $
% \left\{
%   \Psi_P^{(a)}(\tau)
%   \
%   \middle|
%   \
%   0 < a <  P
% \right\}$
It is also related to the colored Jones polynomial for torus link
$T(2,2\, P)$~\cite{KHikami03a}, and 
is a building block for the SU(2) Witten--Reshetikhin--Turaev
invariant for the Seifert manifolds associated with the ADE
singularities~\cite{KHikami05a,KHikami05b}.

%%%
\subsection{Setup of Zwegers}

We first review the results of Zwegers~\cite{Zweg02Thesis}.
We   introduce the Lerch sum as
\begin{equation}
  \label{define_Zwegers_f_P}
  f_P(u, z; \tau)
  =
  \sum_{n \in \mathbb{Z}} 
  \frac{q^{P n^2} \, \E^{4 \pi \I P n z}}{
    1 - q^n \, \E^{2 \pi \I (z-u)}
  } ,
\end{equation}
where $P \in \mathbb{Z}_{>0}$,
and it is related to the level-$k$ of the superconformal algebra as 
$k=P-1$.
% The $q$-series~\eqref{define_Zwegers_f_P}
% is also known as the  higher-level Appell
% function~\cite{SemikTipunTaorm05a}.
Zwegers proved the following formulae;
\begin{gather}
  \begin{gathered}[b]
    f_P(u+1,z;\tau) = f_P(u,z; \tau) ,
    \\[2mm]
    f_P(u,z; \tau)
    -
    q^{-P} \, \E^{-4 \pi \I P u} \,
    f_P(u+\tau , z; \tau)
    =
    \sum_{a=0}^{2P-1}
    q^{- \frac{a^2}{4 P}} \, \E^{-2 \pi \I a u} \,
    \vartheta_{P,a}(z;\tau) ,
    \\[2mm]
    f_P(u,z; \tau+1)
    =
    f_P(u,z; \tau) ,
    \\[2mm]
    f_P(u, z; \tau) - \frac{1}{\tau} \, \E^{2 \pi \I P \frac{u^2 -
        z^2}{\tau}} \,
    f_P \left( {\frac{u}{\tau}},
      \frac{z}{\tau} ; \frac{-1}{\tau} 
    \right)
    =
    \sum_{a=0}^{2P-1}
    M_{P,a}(u;\tau) \, \vartheta_{P,a}(z;\tau) ,
  \end{gathered}
  \label{f_u_transform}
\end{gather}
where
$\vartheta_{P,a}(z;\tau)$ are level-$P$ theta
functions~\eqref{define_vartheta},
and $M_{P,a}(u;\tau)$ has a form of the Mordell integral
\begin{equation}
  \label{define_Zwegers_h}
  M_{P,a}(u;\tau)
  =
  \I \, \E^{- \pi \I \frac{a^2}{2 P} \tau - 2 \pi \I a u} \,
  \int\limits_{\mathbb{R} - \I 0}
  \frac{
    \E^{2 \pi \I P \tau x^2 - 2 \pi
      \left(
        2 P u + a \tau
      \right) x}
  }{
    1 - \E^{2 \pi x}
  } \,
  \mathrm{d} x .
\end{equation}

As a completion of 
$f_P(u,z;\tau)$, we define
$\widehat{f}_P(u,z;\tau)$ by
\begin{equation}
  \widehat{f}_P(u,z;\tau)
  =
  f_P(u,z;\tau)
  - \frac{1}{2} \,
  \sum_{a  \mod 2 P}
  R_{P,a}(u;\tau) \,
  \vartheta_{P,a}(z;\tau),
\end{equation}
where the non-holomorphic partner $R_{P,a}(u;\tau)$ is given by
\begin{multline}
  \label{define_Zwegers_R}
  R_{P,a}(u; \tau)
  \\
  =
  \sum_{
    \substack{
      n \in \mathbb{Z}
      \\
      n \equiv a  \mod 2 P
    }}
  \left[
    \sign\left(n+\frac{1}{2}\right)
    -
    E
    \left(
      \left( n + 2\, P \, \frac{\Im u}{\Im \tau}\right) \,
      \sqrt{\frac{\Im \tau}{P}}
    \right)
  \right] \,
  q^{-\frac{n^2}{4P}} \, \E^{-2 \pi \I n u} .
%  \E^{- \pi \I \frac{n^2}{2 P} \tau - 2 \pi \I n u}
\end{multline}
We then find that
the function $\widehat{f}_P(u,z;\tau)$
has a modular property similar to a 2-variable Jacobi form with weight
$1$;
\begin{equation}
  \begin{gathered}
    \widehat{f}_P(u,z; \tau+1)
    =
    \widehat{f}_P(u, z; \tau),
    \\[2mm]
    \widehat{f}_P \left(
      \frac{u}{\tau} , \frac{z}{\tau} ; - \frac{1}{\tau}
    \right)
    =
    \tau \,
    \E^{2 \pi \I P \frac{z^2 - u^2}{\tau}} \,
    \widehat{f}_P(u,z; \tau) .
  \end{gathered}
\end{equation}

%%%
\subsection{Generalized Mordell Integral}
We study the period function of the weight-$3/2$
vector-valued  modular form
$\{\Psi_P^{(a)}(\tau) ~|~ 0<a<P \}$
defined in~\eqref{Psi_and_chi};
\begin{equation}
  M_P^{(a)}(\tau)
  =
  \int_{0}^{\I \infty}
  \frac{
    \Psi_P^{(a)}(z)
  }{
    \sqrt{\frac{z+\tau}{\I}}
  }  \,
  \mathrm{d} z .
  \label{h_Eichler}
\end{equation}
We substitute~\eqref{Def_Psi} for the above integrand, and apply
an identity~\citep[Lemma~1.18]{Zweg02Thesis}
\begin{equation}
  \int_{-\infty}^\infty
  \frac{
    \E^{\pi \I \tau w^2}
  }{
    w + \I \, r
  } \,
  \mathrm{d} w
  =
  - \pi \, r \,
  \int_0^{\I \infty}
  \frac{\E^{\pi \I r^2 z}}{
    \sqrt{\frac{z+\tau}{\I}}
  } \,
  \mathrm{d} z .
\end{equation}
Then the period function is rewritten as
\begin{equation*}
  M_P^{(a)}(\tau)
  =
  -\frac{\sqrt{2P}}{2 \pi}
  \int\limits_\mathbb{R}
  \mathrm{d} z
  \,
  \E^{\pi \I \tau \frac{z^2}{2 P}} \,
  \sum_{k \in \mathbb{Z}}
  \left(
    \frac{1}{z+\I \, (2 P k +a)}
    -
    \frac{1}{z+\I \, (2 P k -a)}
  \right) .
\end{equation*}
When  we make  use of an infinite series expansion
\begin{equation*}
%   \sum_{n\in \mathbb{Z}} \frac{1}{2x +(2n+1)\I}
%   =
%   \frac{\pi}{2} \frac{\sh(\pi x)}{\ch(\pi x)}
  \tanh\left(\frac{\pi}{2} x \right)
  =
  \frac{4 x}{\pi}
  \sum_{k=1}^\infty
  \frac{
    1}{
    (2k-1)^2+x^2} ,
\end{equation*}
we find that the period function is written
in the form of the generalized
Mordell
integral as
\begin{align}
  &M_P^{(a)}(\tau)
  \nonumber \\
  &  =
  -\frac{1}{2 \sqrt{2 \, P}} 
  \int\limits_\mathbb{R} \mathrm{d} z \,
  \E^{\pi \I \tau \frac{z^2}{2 P}} \,
  \left[
    \tanh \left( 
      \frac{\pi
        \left( z + \I \, (a - P) \right)
      }{2 \, P} 
    \right)
    -
    \tanh \left( 
      \frac{\pi
        \left( z - \I \, (a - P) \right)
      }{2 \, P} 
    \right)
  \right]
  \nonumber
%  \label{Mordell_tanh}
  \\
  & =
  \frac{\I}{\sqrt{2P}}
  \int\limits_\mathbb{R}
  \mathrm{d} z \,
  \E^{\pi \I \tau \frac{z^2}{2 P}} \,
  \frac{
    \sin \left(
      \frac{P-a}{P}\pi \right)
  }{
    \cosh\left(\frac{z}{P}\pi\right)
    +
    \cos \left(
      \frac{P-a}{P} \pi
    \right)
  } .
  \label{Mordell_integral}
\end{align}
This form of the Mordell integral has also been noted
in~\cite{Taorm08a}.
% For our convention we set
% \begin{equation}
%   \begin{aligned}
%     M_P^{(0)}(\tau)
%     & = \I \sqrt{2 \, P},
%     \\[2mm]
%     M_P^{(P)}(\tau) &= 0 .
%   \end{aligned}
%   \label{h_P_rest}
% \end{equation}
% Then the generalized Mordell integral $M_P^{(a)}(\tau)$ is defined for
% $0 \leq a <2 \, P$
% by~\eqref{Mordell_integral} and~\eqref{h_P_rest}.
These 
integrals
%~\eqref{Mordell_integral} and~\eqref{h_P_rest}
have
another form
\begin{equation}
  M_P^{(a)}(\tau)
  =
  \I \, \sqrt{\frac{\I}{\tau}} \,
  \int_{-\infty}^\infty \mathrm{d} x \,
  \E^{-\pi \I \frac{x^2}{2 P \tau}} \,
  \frac{
    \sinh \left( \pi \, \frac{P-a}{P} \, x\right)}{
    \sinh ( \pi \, x)
  } .
\end{equation}
which is known to generate the Ohtsuki-type invariant for torus
link~\cite{KHikami03a}.
Note that the generalized Mordell integral~\eqref{Mordell_integral}
is also used in studies of
the integer partition~\cite{BrinLove07a}.
Expression~\eqref{h_Eichler} of the Mordell integral  as a
period integral easily proves the $S$-transformation formula
\begin{equation}
  \sum_{b=1}^{P-1} \mathbf{S}(P)_{ab} \,
  M_P^{(b)} (\tau)
  =
  \sqrt{\frac{\I}{\tau}} \,
  M_P^{(a)}\left(
    - \frac{1}{\tau}
  \right) ,
\end{equation}
when  we  substitute~\eqref{S_transform} for~\eqref{h_Eichler}.

%%%
Relationship  of $M_P^{(a)}(\tau)$ with
Zwegers'  function~\eqref{define_Zwegers_h} $M_{P,a}(u;\tau)$ is given as follows:
%\citep[Prop.~3.3]{Zweg02Thesis}
From~\eqref{Mordell_integral}, we compute as
\begin{align*}
  M_P^{(a)}(\tau)
  &=
  \frac{\E^{- \pi \I \tau \frac{(a-P)^2}{2 P}}}{2 \sqrt{2 \, P}} \,
  \int\limits_\mathbb{R} \mathrm{d} w \,
  \tanh\left( \frac{\pi w}{2  P}\right) \,
  \E^{\pi \I \tau \frac{w^2}{2 P}} \,
  \left(
    \E^{\pi \tau \frac{P-a}{P}w}
    -    \E^{-\pi \tau \frac{P-a}{P}w}
  \right)
  \\
%   &=
%   \frac{\E^{- \pi \I \tau \frac{(a-P)^2}{2 P}}}{ \sqrt{2 \, P}} \,
%   \int\limits_\mathbb{R} \mathrm{d} w \,
%   \tanh\left( \frac{\pi w}{2  P}\right) \,
%   \E^{\pi \I \tau \frac{w^2}{2 P} +
%     \pi \tau \frac{P-a}{P}w}
%   \\
%   & =
%   \frac{\E^{- \pi \I \tau \frac{(a-P)^2}{2 P}}}{ \sqrt{2 \, P}} \,
%   \int\limits_{\mathbb{R}- \I 0}
%   \mathrm{d} z \,
%   \coth\left( \frac{\pi z}{2  P}\right) \,
%   \E^{\pi \I \tau \frac{(z+P\I)^2}{2 P} +
%     \pi \tau \frac{P-a}{P} (z+P\I)} .
  & =
  \frac{\E^{- \pi \I \frac{a^2}{2 P} \tau}}{\sqrt{2 \, P}}
  \left[
    \int\limits_{\mathbb{R}+ \I 0}
    \frac{\E^{\pi \I \tau \frac{z^2}{2P} - \pi a \tau \frac{z}{P}}}{
      1 - \E^{- \pi \frac{z}{P}}} \,
    \mathrm{d} z
    -
    \int\limits_{\mathbb{R}- \I 0}
    \frac{\E^{\pi \I \tau \frac{z^2}{2P} - \pi a \tau \frac{z}{P}}}{
      1 - \E^{\pi \frac{z}{P}}} \,
    \mathrm{d} z
  \right] .
\end{align*}
Thus  we obtain
\begin{align}
  M_P^{(a)}(\tau)
  =
  \I \, \sqrt{2 \, P} \,
  \left[
    M_{P,a}(0;\tau) -   M_{P,-a}(0;\tau)     \right] .
%   & =
%   \begin{cases}
%     h_P^{(a)}(\tau) & \text{for $0<a<P$, $P<a<2\,P$}
%     \\
%     \I \, \sqrt{2 \, P}
%     & \text{for $a=0$}
%     \\
%     0 &
%     \text{for $a=P$}    
%   \end{cases}
  \label{Mordell_and_Zwegers}
\end{align}
%for $0 \leq a < 2 \, P$.

\subsection{Non-Holomorphic Function}
We define a non-holomorphic function
\begin{gather}
  \label{define_R_P_a}
  R_P^{(a)}(\tau)
  =
  \int_{- \overline{\tau}}^{\I \infty}
  \frac{
    \Psi_P^{(a)}(z)
  }{
    \sqrt{\frac{z+\tau}{\I}}
  } \,
  \mathrm{d} z ,
\end{gather}
as a generalization of~\eqref{R0_integral}.
This integral expression shows that
\begin{equation}
  \label{differential_R_P_a}
  \frac{\partial}{\partial \overline{\tau}} \,
  R_P^{(a)}(\tau)
  =
  \frac{
    \Psi_P^{(a)}(- \overline{\tau})
  }{
    \sqrt{2 \, \Im \tau}
  } .
\end{equation}
% which shows
% \begin{equation}
%   \frac{\partial}{\partial \tau}  \sqrt{\Im \tau} 
%   \frac{\partial}{\partial \overline{\tau}} \,
%   R_P^{(a)}(\tau)=0
% \end{equation}
By integrating  each summand of $\Psi_P^{(a)}(z)$ after
substitution of~\eqref{Def_Psi} into~\eqref{define_R_P_a},
we find
% \begin{align*}
%   R_P^{(a)}(\tau)
%   = \frac{1}{2} \sum_{n \in \mathbb{Z}}
%   n \,  \psi_{2P}^{(a)}(n)  \,
%   \int_{- \overline{\tau}}^{\I \infty}
%   \frac{
%     \E^{ \pi \I z \frac{n^2}{2 P}}
%   }{
%     \sqrt{\frac{z+\tau}{\I}}
%   } \,
%   \mathrm{d} z
% \end{align*}
% which gives
\begin{equation}
  R_P^{(a)}(\tau) =
  \I \, \sqrt{2 \, P} \,
%  \E^{2 \pi \I \frac{a b}{2 P}} \,
  R_{P,a} \left(
    0; \tau \right) ,
%    \frac{b}{2 P} ; \tau \right)
\end{equation}
%where $b \in \mathbb{Z}$.
where $R_{P,a}(z;\tau)$ is   Zwegers' $R$-function
%\citep[Definition~3.4]{Zweg02Thesis} is
defined
in~\eqref{define_Zwegers_R}.

Substituting  the modular transformation law~\eqref{S_transform}
for $\Psi_P^{(a)}(z)$ in~\eqref{define_R_P_a},
we find
the  transformation law for $R_P^{(a)}(\tau)$  as follows;
\begin{equation}
  \begin{gathered}
    \sqrt{\frac{\I}{\tau}} \,
    \sum_{b=1}^{P-1}
    \mathbf{S}(P)_{ab}
    \,
    R_P^{(b)}\left( - \frac{1}{\tau} \right)
    +
    R_P^{(a)}(\tau)
    =
    M_P^{(a)}(\tau) ,
    \\[2mm]
    R_P^{(a)}(\tau+1)
    =
    \E^{- \pi \I \frac{a^2}{2 P}} \, R_P^{(a)}(\tau) ,
  \end{gathered}
\end{equation}
where $M_P^{(a)}(\tau)$ is the Mordell integral defined
in~\eqref{h_Eichler}.
Using  the level-$P$ $\vartheta$-function~\eqref{define_vartheta},
the $S$-transformation formula is rewritten as
\begin{multline}
  \sum_{a=1}^{P-1} M_P^{(a)}(\tau) \,
  \left(
    \vartheta_{P,a}
    -
    \vartheta_{P,-a}
  \right) (z;\tau)
  =
  \left[
    \sum_{a=1}^{P-1} R_P^{(a)}(\tau) \,
    \left(
      \vartheta_{P,a}
      -
      \vartheta_{P,-a}
    \right) (z;\tau)
  \right]
  \\
  -\frac{1}{\tau} \, \E^{-2 \pi \I \frac{P}{\tau} z^2} \,
  \left[
    \sum_{a=1}^{P-1}
    R_P^{(a)}\left( -\frac{1}{\tau} \right) \,
    \left(
      \vartheta_{P,a}
      -
      \vartheta_{P,-a}
    \right) \left(\frac{z}{\tau}; - \frac{1}{\tau}
    \right)
  \right] .
  \label{general_R_S}
\end{multline}

%%%
\subsection{Modular Transformation of Massless Characters}

%\subsection{Higher Level: reformulation}
The massless character~\eqref{level-k_ch_0} 
is identified with an anti-symmetric part of
the Lerch
sum~\eqref{define_Zwegers_f_P}.
Using~\eqref{f_u_transform} and~\eqref{Mordell_and_Zwegers}, we find
% We substitute $u=0$ for
% \eqref{f_u_transform}, and 
% also 
% $u=-\frac{1}{2 P}$ 
% for \eqref{f_u_transform}
% replacing $z$ with $z-\frac{1}{2 P}$.
% Adding these two identities,
% we get by use of~\eqref{Mordell_and_Zwegers}
\begin{multline*}
  \frac{1}{\I \, \sqrt{2 \, P}}
  \sum_{a=1}^{P-1}
  M_P^{(a)}(\tau) \,
  \left[
    \vartheta_{P,a}(z;\tau)
    -
    \vartheta_{P,-a}(z;\tau)
  \right]
  \\
  =
  \left[
    \sum_{n \in \mathbb{Z}}
    q^{P n^2} \, \E^{4 \pi \I P n z} \,
    \frac{
      1 + q^n \, \E^{2 \pi \I z}
    }{
      1 - q^n \, \E^{2 \pi \I z}
    }
  \right]
  \\
  - \frac{1}{\tau} \,
  \E^{- 2 \pi \I  P \frac{z^2}{\tau}}
  \,
  \left[
    \sum_{n\in \mathbb{Z}} 
    \widetilde{q}^{ P n^2} \,
    \E^{4 \pi \I P n \frac{z}{\tau}} \,
    \frac{
      1 + \widetilde{q}^{ n } \, \E^{2 \pi \I \frac{z}{\tau}}
    }{
      1 - \widetilde{q}^{ n } \, \E^{2 \pi \I \frac{z}{\tau}}
    } 
  \right] ,
%  \label{general_Lerch_S}
\end{multline*}
where $\widetilde{q}=\E^{-2 \pi \I/\tau}$.
We thus obtain a formula
which is a generalization of~\eqref{k1_character_S};
\begin{align}
  F_P(z;\tau)
  +
  \E^{-2 \pi \I (P-2) \frac{z^2}{\tau}} \,
  F_P \left(
    \frac{z}{\tau} ; 
    - \frac{1}{\tau}
  \right)
  & =
  \frac{1}{\I \sqrt{2 \, P}} \,
  \sum_{a=1}^{P-1}
  \frac{M_P^{(a)}(\tau)}{\eta(\tau)} \cdot
  \frac{
    \vartheta_{P,a}
    -
    \vartheta_{P,-a}
  }{
    \vartheta_{2,1}
    -
    \vartheta_{2,-1}
  }(z; \tau)
  \label{general_k_F}
  \\
  & =
  \frac{1}{\I \sqrt{2 \, P}} \,
  \sum_{a=1}^{P-1}
  \frac{M_P^{(a)}(\tau)}{\eta(\tau)} \,
  \chi_{P-2,\frac{a-1}{2}}(z;\tau) ,
  \nonumber
\end{align}
where we have defined
\begin{equation}
  F_P(z;\tau)
  =
%   \frac{1}{\eta(\tau)} \cdot
%   \frac{
%     q^{-\frac{1}{8}} \, \E^{-2 \pi \I z}
%   }{
%     \left(
%       q, \E^{-4 \pi \I z}, q \, \E^{4 \pi \I z}
%     \right)_\infty
%   }
  \frac{
    \I}{
    \eta(\tau) \, \theta_{11}(2 \, z; \tau)
  }
  \,
  \sum_{n \in \mathbb{Z}}
  q^{P n^2} \, \E^{4 \pi \I P n z} \,
  \frac{
    1+ q^n \, \E^{2 \pi \I z}
  }{
    1- q^n \, \E^{2 \pi \I z}
  } .
\end{equation}
We see
that~\eqref{general_k_F} with $P=2$ coincides
with~\eqref{k1_character_S},
and the massless character~\eqref{level-k_ch_0} is related to
$F_P(z;\tau)$ as
\begin{equation}
  \ch_{k,\frac{k}{4},\ell=0}^{\widetilde{R}}(z;\tau)
  =
  \frac{
    \left[ \theta_{11}(z;\tau) \right]^2
  }{
    \left[ \eta(\tau) \right]^3
  } \,
  F_{k+1}(z;\tau)  .
\end{equation}
% We  indeed have
% \begin{align}
%   F_2(z;\tau)
%   &=
%   H(z;\tau)
% \end{align}

%%%%
\subsection{Harmonic Maass Form}

Looking at 
the $S$-transformation~\eqref{general_k_F} of the massless character,
we see that the Mordell integrals can be
compensated  with those from~\eqref{general_R_S}.
By setting $C_P(z;\tau)$ to be the massless
character~\eqref{level-k_ch_0}
 for notational simplicity,
\begin{align}
  C_P(z;\tau)
  & =
  \ch_{k=P-1,h=\frac{P-1}{4},\ell=0}^{\widetilde{R}}(z;\tau) 
  \nonumber
  \\
  & =
  \frac{
    \left[\theta_{11}(z;\tau) \right]^2}{
    \left[ \eta(\tau) \right]^3
  }
  \,
  \frac{
    \I}{
    \theta_{11}(2 \, z; \tau)
  }
  \,
  \sum_{n \in \mathbb{Z}}
  q^{P n^2} \, \E^{4 \pi \I P n z} \,
  \frac{
    1+ q^n \, \E^{2 \pi \I z}
  }{
    1- q^n \, \E^{2 \pi \I z}
  } ,
  \label{define_C_P}
\end{align}
we introduce a (real analytic) Jacobi as
\begin{equation}
  \label{general_Maass_pre}
  \widehat{C}_P(z;\tau)
  =
  C_P(z;\tau)
  -
  \frac{1}{\I \sqrt{2 \, P}} \,
  \sum_{a=1}^{P-1}
  R_P^{(a)}(\tau)
  \, B_P^{(a)}(z;\tau) .
%   =
%   C_P(z;\tau)
%   -
%   \frac{1}{\I \sqrt{2 \, P}}
%   \,
%   \left[
%     \frac{\theta_{11}(z;\tau)}{
%       \eta(\tau)}
%   \right]^2
%   \sum_{a=1}^{P-1}
%   \frac{
%     R_P^{(a)}(\tau)}{
%     \eta(\tau)
%   } \,
%   \frac{
%     \vartheta_{P,a}
%     -
%     \vartheta_{P,-a}
%   }{
%     \vartheta_{2,1}
%     -
%     \vartheta_{2,-1}
%   }(z; \tau) .
\end{equation}
Here the basis function 
$B_P^{(a)}(z;\tau)$ is defined as
\begin{equation}
  B_P^{(a)}(z;\tau)
  =
  \frac{
    \left[ \theta_{11}(z;\tau) \right]^2
  }{
    \left[
      \eta(\tau)
    \right]^3
  } \cdot
  \frac{
    \vartheta_{P,a} - \vartheta_{P,-a}
  }{
    \vartheta_{2,1} - \vartheta_{2,-1}
  }(z; \tau) 
  =
  \frac{
    \left[ \theta_{11}(z;\tau) \right]^2
  }{
    \left[
      \eta(\tau)
    \right]^3
  } \,
  \chi_{P-2,\frac{a-1}{2}}(z;\tau) .
  \label{basis-function}
\end{equation}
%which  corresponds to  the massive character
%\begin{equation*}
%  B_P^{(a)}(z;\tau)
 % =
 % (-1)^{a+1} \, q^{\frac{a^2}{4 P}} \,
 % \lim_{h \searrow \frac{k}{4}}
 % \ch^{\widetilde{R}}_{k=P-1, h, \ell=\frac{a}{2}}(z;\tau) .
%\end{equation*}
The modular  transformation formulae
of $\widehat{C}_P(z;\tau)$
are summarized as follows;
\begin{equation}
  \begin{gathered}
    \widehat{C}_P(z;\tau)
    =
    \E^{-2 \pi \I(P-1) \frac{z^2}{\tau}} \,
    \widehat{C}_P\left(\frac{z}{\tau};-\frac{1}{\tau}\right) ,
    \\[2mm]
    \widehat{C}_P(z;\tau+1)
    =
%     \widehat{C}_P(z;\tau)
%     \\[2mm]
% %
    \widehat{C}_P(z+1 ;\tau)
    =
    \widehat{C}_P(z ;\tau) ,
    \\[2mm]
    \widehat{C}_P(z+\tau ;\tau)
    =
    q^{-(P-1)} \, \E^{-4 \pi \I(P-1) z} \,
    \widehat{C}_P(z ;\tau) .
  \end{gathered}
\end{equation}
We note that the Fourier expansions have the form 
\begin{equation}
  \label{C_X_expand}
  \begin{aligned}
    C_P(z;\tau)
    & =
    1+ \left( \E^{\pi \I z} - \E^{- \pi \I z} \right)^2 \,
    \left( \E^{2 \pi \I z} + \E^{- 2\pi \I z} \right) \, q
    +
    \cdots ,
    \\[2mm]
    B_P^{(a)}(z;\tau)
    & =
    - \left( \E^{\pi \I z} - \E^{- \pi \I z} \right)^2 \,
    I_{\frac{a-1}{2}}(z) \, q^{\frac{a^2}{4 P}}
    +
    \cdots ,
  \end{aligned}
\end{equation}
where $I_{\frac{a}{2}}(z)$ is associated with the Chebyshev polynomial
of the second kind;
\begin{equation*}
  I_{\frac{a}{2}}(z) =
  \sum_{n = - \frac{a}{2}}^{\frac{a}{2}} \E^{4 \pi \I n z} .
\end{equation*}

To construct an eigenfunction of the differential
operator~\eqref{differential_Maass} at a higher level,
we introduce a holomorphic function
\begin{equation}
  \label{define_M_P}
  {H}_P^{(a)}(z; \tau)
  =
  \frac{
    \Wron\left[
      B_P^{(a)}\to {C}_P
    \right]
  }{
    \Wron\left[
      B_P^{(1)}, \dots,
      B_P^{(P-1)}
    \right]
  }(z;\tau) ,
\end{equation}
whose completion is  given by
\begin{equation}
  \label{define_hat_M_P}
  \widehat{H}_P^{(a)}(z; \tau)
  =
  \frac{
    \Wron\left[
      B_P^{(a)}\to \widehat{C}_P
    \right]
  }{
    \Wron\left[
      B_P^{(1)}, \dots,
      B_P^{(P-1)}
    \right]
  }(z;\tau) .
\end{equation}
Here
$\Wron \left[ \phi_1,\dots, \phi_n \right](z;\tau)$ 
denotes the Wronskian
with respect to $z$;
\begin{equation*}
  \Wron \left[ \phi_1,\dots,\phi_n \right](z;\tau)
  =
  \left\Norm
    \begin{matrix}
      \phi_1(z;\tau) & \dots & \phi_n(z;\tau)
      \\
      \vdots & \ddots & \vdots 
      \\
      \frac{
        \mathrm{d}^{n-1}  \phi_1
      }{
        \mathrm{d} z^{n-1}
      }
      (z;\tau)
      & \dots &
      \frac{
        \mathrm{d}^{n-1}  \phi_n
      }{
        \mathrm{d} z^{n-1}
      }
      (z;\tau)
    \end{matrix}
  \right\Norm 
\end{equation*}
and 
$\Wron\left[
  B_P^{(a)} \to \widehat{C}_P
\right](z;\tau)$ means the Wronskian
$\Wron\left[B_P^{(1)}, \dots, B_P^{(P-1)} \right](z;\tau)$
with
$B_P^{(a)}(z;\tau)$ replaced by $\widehat{C}_P(z;\tau)$.
From~\eqref{general_Maass_pre}
and~\eqref{differential_R_P_a}, we have
\begin{align}
  \frac{\partial }{\partial \overline{\tau}}
  \widehat{H}_P^{(a)}(z;\tau)
  & =
  \frac{\I}{\sqrt{2 \, P}} \cdot
  \frac{\partial R_P^{(a)}(\tau)}{\partial \overline{\tau}}
  \nonumber
  \\
  & =
  \frac{\I}{\sqrt{2 \, P}} \cdot
  \frac{
    \Psi_P^{(a)}(- \overline{\tau})
  }{
    \sqrt{2 \, \Im \tau}
  } .
\end{align}
As a result, we have
\begin{equation*}
  \left( \Im \tau \right)^{\frac{3}{2}}
  \frac{\partial}{\partial \tau}  \sqrt{\Im \tau} 
  \frac{\partial}{\partial \overline{\tau}} \,
  \widehat{H}_P^{(a)}(z;\tau)
  = 0 ,
\end{equation*}
which,
with $\tau=u+\I \, v$, reduces to 
\begin{equation}
%   \Delta_{\frac{1}{2}} \widehat{H}_P^{(a)}(z;\tau)
%   =
  \left[
    -v^2 \,
    \left( \frac{\partial^2}{\partial u^2}
      +
      \frac{\partial^2}{\partial v^2}
    \right)
    +\frac{\I \, v}{2} \,
    \left(
      \frac{\partial}{\partial u}
      + \I \,
      \frac{\partial}{\partial v}
    \right)
  \right] \widehat{H}_P^{(a)}(z;\tau)
  = 0 .
\end{equation}

We can check the modular transformation formulae as
\begin{equation}
  \begin{gathered}
    \widehat{H}_P^{(a)}(z;\tau)
    =
    -
    \sqrt{\frac{\I}{\tau}} \,
    \sum_{b=1}^{P-1} \mathbf{S}(P)_{ab} \,
    \widehat{H}_P^{(b)}\left(
      \frac{z}{\tau} ; - \frac{1}{\tau}
    \right) ,
    \\[2mm]
    \widehat{H}_P^{(a)}(z;\tau+1) =
    \E^{-\frac{a^2}{2 P} \pi \I} \,
    \widehat{H}_P^{(a)}(z;\tau) ,
    \\[2mm]
    \widehat{H}_P^{(a)}(z+1;\tau)
    =
    \widehat{H}_P^{(a)}(z+\tau;\tau)
    =
    \widehat{H}_P^{(a)}(z;\tau)  .   
  \end{gathered}
\end{equation}
Thus recalling~\cite{Shimura73a},
we find that
%$\widehat{H}_P^{(a)}(z;\tau)$ is
these are
the harmonic Maass form with
weight $1/2$ on $\Gamma(4P)$.

%%%%
\subsection{Character Decomposition}
We  define an element of
$(P-1)\times(P-1)$ matrix $\mathbf{B}_P(\boldsymbol{w};\tau)$
by
\begin{equation}
  \left(
    \mathbf{B}_P(\boldsymbol{w};\tau)
  \right)_{ab} =
  B_P^{(b)}(w_a;\tau) ,
\end{equation}
where $\left\{ w_1,  w_2, \dots, w_{P-1} \right\}$ are coordinates on the torus.
As a higher-level generalization of~\eqref{define_lowest_T},
we define
$J_P(z;w_1,\dots,w_{P-1};\tau)$ by
\begin{align}
  &J_P(z; w_1, \dots, w_{P-1}; \tau)
  \nonumber
  \\
  &=
  \widehat{C}_P(z;\tau) -
  \sum_{a=1}^{P-1} \sum_{b=1}^{P-1}
  \left(
    \mathbf{B}_P(\boldsymbol{w};\tau)^{-1} \right)_{ab} \,
  \widehat{C}_P(w_b;\tau)  \,
  B_P^{(a)}(z;\tau)
  \label{Maass_sum}
  \\
  & =
  {C}_P(z;\tau) -
  \sum_{a=1}^{P-1} \sum_{b=1}^{P-1}
  \left(
    \mathbf{B}_P(\boldsymbol{w};\tau)^{-1} \right)_{ab} \,
  {C}_P(w_b;\tau)  \,
  B_P^{(a)}(z;\tau) ,
  \label{Maass_sum2}
  %  \label{Maass_sum_holomorphic}
\end{align}
where the second equality follows from the $z$-independence of the function $R^{(a)}_P(\tau)$~\eqref{general_Maass_pre}.
The function $J_P(z; w_1,\dots,w_{P-1};\tau)$  has a holomorphic
$q$-series as seen
from the second equality, and 
it has a good
behavior under  modular transformations.
%due to the definition~\eqref{Maass_sum}.
Explicitly we have
\begin{equation}
  \label{transform_T_P}
  \begin{gathered}
    J_P(z;w_1,\dots,w_{P-1};\tau)
    =
    \E^{-2 \pi \I (P-1)\frac{z^2}{\tau}} \,
    J_P\left(
      \frac{z}{\tau}; \frac{w_1}{\tau}, \dots, \frac{w_{P-1}}{\tau};
      -\frac{1}{\tau}
    \right) ,
    \\[2mm]
    \begin{aligned}
      J_P(z+1;w_1,\dots,w_{P-1};\tau)
      & =
      J_P(z;w_1,\dots,w_a+1,\dots,w_{P-1};\tau)
      \\
      & =
      J_P(z;w_1,\dots,w_a+\tau,\dots,w_{P-1};\tau)
      \\
      & =
      J_P(z;w_1,\dots,w_{P-1};\tau+1)
      \\
      &    =
      J_P(z;w_1,\dots,w_{P-1};\tau) ,
    \end{aligned} 
    \\[2mm]
    J_P(z+ \tau;w_1,\dots,w_{P-1};\tau)
    =
    q^{-(P-1)} \, \E^{-4 \pi \I (P-1) z} \,
    J_P(z;w_1,\dots,w_{P-1};\tau) .
  \end{gathered}
\end{equation}
By construction, the function $J_P(z; w_1,\dots,w_{P-1};\tau)$
vanishes at $z=w_a$ for $a=1,\dots,P-1$,
\begin{equation}
  J_P(w_a; w_1,\dots, w_{P-1};\tau) = 0 ,
\end{equation}
and we also have
\begin{equation}
  J_P(0; w_1,\dots, w_{P-1};\tau) = 1.
\end{equation}

\subsubsection{Degenerate Case}
We consider the completely degenerate configuration,
$w_1=w_2=\dots=w_{P-1}$;
we set
\begin{equation}
  J_P(z;w;\tau)
  =
  \lim_{\substack{
      w_a \to w
      \\
      \forall a}}
  J_P(z;w_1,\dots,w_{P-1};\tau)  .
\end{equation}
We then obtain from~\eqref{Maass_sum}
\begin{align}
  \label{define_T_P_degenerate}
  J_P(z;w;\tau)
  & =
  \widehat{C}_P(z;\tau) -
  \sum_{a=1}^{P-1} \widehat{H}_P^{(a)}(w;\tau) \, B_P^{(a)}(z;\tau)
  \\
  & =
  {C}_P(z;\tau) -
  \sum_{a=1}^{P-1} {H}_P^{(a)}(w;\tau) \, B_P^{(a)}(z;\tau) ,
\end{align}
where  $\widehat{H}_P^{(a)}(z;\tau)$ is the harmonic Maass form
of~\eqref{define_hat_M_P},
and $H_P^{(a)}(z;\tau)$ is its holomorphic part~\eqref{define_M_P}. 
The second line above gives a decomposition of a massless character
$C_P(z;\tau)$  into a 
Jacobi form $J_P(z;w;\tau)$ and a sum 
over basis functions $B_P^{(a)}(z;\tau)$.  
Note that
$J_P(z;w;\tau)$ has $2(P-1)$-th order zero at $z=w$.
Its transformation formulae are given as
\begin{equation}
  \begin{gathered}
    J_P(z;w;\tau)
    =
    \E^{-2 \pi \I (P-1) \frac{z^2}{\tau}} \,
    J_P\left(
      \frac{z}{\tau} ; \frac{w}{\tau} ; \frac{-1}{\tau}
    \right) ,
    \\[2mm]
    \begin{aligned}
      J_P(z+1;w;\tau) 
      & = J_P(z;w+1;\tau)
      \\
      & = J_P(z;w+\tau;\tau)
      \\
      & =
      J_P(z;w; \tau+1)=J_P(z;w;\tau) ,
    \end{aligned}
    \\[2mm]
    J_P(z+\tau; w;\tau)
    =
    q^{-(P-1)} \, \E^{-4 \pi \I (P-1) z} \,
    J_P(z;w;\tau) .
  \end{gathered}
\end{equation}
When we set $w$ to special values
$\left\{
  \frac{1}{2}, \frac{1+\tau}{2}, \frac{\tau}{2}
\right\}$, these formulas give
the following identifications;
\begin{equation}
  \label{T_P_degenerate}
    J_P \left( z;\frac{1}{2}; \tau \right)
    =
    \left(
      \frac{\theta_{10}(z;\tau)}{
        \theta_{10}(0;\tau)}
    \right)^{\hskip-1mm 2(P-1)}\hskip-10mm , \hskip3mm 
    J_P \left( z;\frac{1+\tau}{2}; \tau \right)
   =
    \left(
      \frac{\theta_{00}(z;\tau)}{
        \theta_{00}(0;\tau)}
    \right)^{\hskip-1mm 2(P-1)} \hskip-10mm , \hskip3mm 
    J_P \left( z;\frac{\tau}{2}; \tau \right)
   =
    \left(
      \frac{\theta_{01}(z;\tau)}{
        \theta_{01}(0;\tau)}
    \right)^{\hskip-1mm 2(P-1)} \hskip-10mm .
\end{equation}
Hence we have 
\begin{eqnarray}
  \hskip-3mm   \ch_{k=P-1,h=\frac{P-1}{4},\ell=0}^{\widetilde{R}}(z;\tau)     &=&
     \left(
      \frac{\theta_{10}(z;\tau)}{
        \theta_{10}(0;\tau)}
    \right)^{\hskip-1mm 2(P-1)}    +\sum_{a=1}^{P-1} H_P^{(a)}
    \left({1\over 2};\tau \right)  \,B_P^{(a)}(z;\tau)
    \label{character-decomposition1}
    \\
    &=&  \left(
      \frac{\theta_{00}(z;\tau)}{
        \theta_{00}(0;\tau)}
    \right)^{\hskip-1mm 2(P-1)}    
      \hskip-2mm +\sum_{a=1}^{P-1} H_P^{(a)} \left({1+\tau\over 2};\tau\right) \,
    B_P^{(a)}(z;\tau) \label{character-decomposition2}
    \\
    &=&
    \left(
      \frac{\theta_{01}(z;\tau)}{
        \theta_{01}(0;\tau)}
    \right)^{\hskip-1mm 2(P-1)}    +
    \sum_{a=1}^{P-1} H_P^{(a)} \left({\tau\over 2};\tau\right) \,
    B_P^{(a)}(z;\tau). \label{character-decomposition3}
\end{eqnarray}
% \begin{eqnarray}
%   &&C_P(z;\tau)=J_P(z;{1\over 2};\tau) \hskip0.3cm +\hskip0.5cm \sum H_P^{(a)}({1\over 2};\tau)B_P^{(a)}(z;\tau),
%   \nonumber\\
%   &&C_P(z;\tau)=J_P(z;{1+\tau\over 2};\tau)+\sum H_P^{(a)}({1+\tau\over 2};\tau)B_P^{(a)}(z;\tau),\\
%   \label{character-decomposition}
%   &&C_P(z;\tau)=J_P(z;{\tau\over 2};\tau) \hskip0.3cm +\hskip0.5cm \sum H_P^{(a)}({\tau\over 2};\tau)B_P^{(a)}(z;\tau).
%   \nonumber
% \end{eqnarray}

\subsection{Elliptic Genera}      
Character decomposition above can be  used to rewrite the
elliptic genera of hyper-K\"ahler manifolds $X_k$ with
$\dim_{\mathbb{C}}X_k=2k$,
$k\ge 1$.

We note that the elliptic genera for hyper-K\"ahler manifolds $X_k$ is in general given by a sum of symmetric polynomials of  $\left\{
  \left( \frac{\theta_{10}(z;\tau)}{\theta_{10}(0;\tau)} \right)^2,
  \left( \frac{\theta_{00}(z;\tau)}{\theta_{00}(0;\tau)} \right)^2,
  \left( \frac{\theta_{01}(z;\tau)}{\theta_{01}(0;\tau)} \right)^2
\right\}$ with order $k$. 
In the case of a polynomial of the following form 
\begin{equation}
  \label{define_Z_X_k}
  Z^{(1)}_{X_k}(z;\tau)
  =
  c_k\,2^{2k} \,
  \left[
    \left(
      \frac{\theta_{10}(z;\tau)}{\theta_{10}(0;\tau)}
    \right)^{2k}
    +
    \left(
      \frac{\theta_{00}(z;\tau)}{\theta_{00}(0;\tau)}
    \right)^{2k}
    +
    \left(
      \frac{\theta_{01}(z;\tau)}{\theta_{01}(0;\tau)}
    \right)^{2k}
  \right] ,
\end{equation}
we can use the character decomposition of the completely degenerate
configuration 
(\ref{character-decomposition1}),
(\ref{character-decomposition2}),
(\ref{character-decomposition3}). 
Here we introduced a factor $2^{2k}$ for convenience and the  parameter $c_k \in \mathbb{Z}_{>0}$ is to be fixed later so that $Z^{(1)}_{X_k}$ contains the identity representation 
with multiplicity $1$ in the NS sector.

In the case of a polynomial of the form 
\begin{equation*}
  %J_{k+1}(z; \boldsymbol{w}_{k_2,k_3,k_4};\tau)
  %=
  \left[
    \left( \frac{\theta_{10}(z;\tau)}{\theta_{10}(0;\tau)} \right)^{2 k_2}
    \left( \frac{\theta_{00}(z;\tau)}{\theta_{00}(0;\tau)} \right)^{2 k_3}
    \left( \frac{\theta_{01}(z;\tau)}{\theta_{01}(0;\tau)} \right)^{2 k_4}
    +
    \text{permutations}
  \right] 
\end{equation*}
one can use the character decomposition ({\ref{Maass_sum2}) corresponding to the choice of parameters 
$w_1,\cdots,w_{P-1}$ 
\begin{equation*}
  \boldsymbol{w}_{k_2,k_3,k_4}
  =
  \left\{  \begin{array}{cc}  
  w_1, \dots, w_k 
    ~|~
       &  \#\left(w_a=\frac{1}{2}\right)=k_2,
      \#\left(w_a=\frac{1+\tau}{2}\right)=k_3,
      \#\left(w_a=\frac{\tau}{2}\right)=k_4,
      \\
 \null &     k_2+k_3+k_4=k.
    \end{array}
  \right\} .
\end{equation*}
In fact the Jacobi form corresponding to the vector $\boldsymbol{w}_{k_2,k_3,k_4}$ is given by
\begin{equation*}
  J_{k+1}(z; \boldsymbol{w}_{k_2,k_3,k_4};\tau)
  = \left( \frac{\theta_{10}(z;\tau)}{\theta_{10}(0;\tau)} \right)^{2 k_2}
  \left( \frac{\theta_{00}(z;\tau)}{\theta_{00}(0;\tau)} \right)^{2 k_3}
  \left( \frac{\theta_{01}(z;\tau)}{\theta_{01}(0;\tau)} \right)^{2 k_4}.
\end{equation*}

Let us first consider the completely degenerate case.
Using~\eqref{character-decomposition1}--\eqref{character-decomposition3}
and~\eqref{basis-function}, 
we obtain 
\begin{multline*}
  c_k^{~-1} \, 2^{-2k}
  Z_{X_k}^{(1)}(z;\tau)
  =
  3 \ch_{k,h=\frac{k}{4},\ell=0}^{\widetilde{R}}(z;\tau)
  \\
  -
  \sum_{a=1}^k
  \sum_{
    w
    \in
    \left\{
      \frac{1}{2},
      \frac{1+\tau}{2},
      \frac{\tau}{2}
    \right\}
  }  H_{k+1}^{(a)}(w;\tau)   \,
  \frac{
    \left[ \theta_{11}(z;\tau) \right]^2
  }{
    \left[ \eta(\tau) \right]^3
  } \,
  \chi_{k-1,\frac{a-1}{2}}(z;\tau).
\end{multline*}
It turns out that the coefficients of
lowest powers of
the Fourier expansion of 
$H_P^{(a)}(w;\tau)$ 
with 
$\tau \in
\left\{
  \frac{1}{2}, \frac{1+\tau}{2} , \frac{\tau}{2}
\right\}$
are conveniently computed from the expansion of Jacobi forms ~\eqref{T_P_degenerate}.
Namely by use of~\eqref{C_X_expand} and~\eqref{theta_power_expand}
with a help of an identity
of the Chebyshev polynomial 
\begin{equation*}
  \left( \E^{\pi \I z} - \E^{- \pi \I z} \right)^{2n}
  =
  (-1)^n
  \sum_{k=0}^n
  (-1)^k \frac{2 \, (k+1)}{n+k+2} \,
  \begin{pmatrix}
    2 \, n+1 \\
    n-k
  \end{pmatrix} \,
  I_{\frac{k}{2}}(z) ,
\end{equation*}
we find that
\begin{equation}
  \label{M_P_Fourier}
  \begin{aligned}
    H_P^{(a)} \left( \frac{1}{2} ; \tau \right)
    & =
    q^{-\frac{a^2}{4 P}} \,
    \left(
      \alpha_P^{(a)} + \mathcal{O}(q)
    \right) ,
    \\[2mm]
    H_P^{(a)} \left( \frac{1+\tau}{2} ; \tau \right)
    & =
    q^{- \frac{a^2}{4 P} + \frac{a}{2}} \,
    \left(
      \beta_{P-1,P-1-a} + \mathcal{O}(q^{\frac{1}{2}})
    \right) ,
    \\[2mm]
    H_P^{(a)} \left( \frac{\tau}{2} ; \tau \right)
    & =
    q^{- \frac{a^2}{4 P} + \frac{a}{2}} \,
    \left(
      (-1)^a \cdot
      \beta_{P-1,P-1-a} + \mathcal{O}(q^{\frac{1}{2}})
    \right) .
  \end{aligned}
\end{equation}
Here we have defined  $\alpha_P^{(a)}$\,$(a=1,2,\dots,P-1)$ 
and
$\beta_{n,k}$ by
\begin{equation}
  \begin{gathered}
    \alpha_P^{(a)}
    =
    \sum_{m=a}^{P-1}
    \frac{    (-1)^{m+a}
    }{2^{2m-1}} \,
    \frac{
      a
    }{ m+ a  } \,
    \begin{pmatrix}
      P-1 \\
      m
    \end{pmatrix} 
    \,
    \begin{pmatrix}
      2 \, m -1 \\
      m-a
    \end{pmatrix}
    ,
    \\[2mm]
    \beta_{n,k}
    =
    \frac{2 (k+1)}{
      n+k+2} \,
    \begin{pmatrix}
      2\, n+1 \\
      n-k
    \end{pmatrix}  .
  \end{gathered}
\end{equation}
These results agree with Fourier coefficients of $H_P^{(a)}(z;\tau)$
computed directly from~\eqref{define_M_P}
in Appendix~\ref{sec:Fourier_Maass}.

As we see from the Appendix, the $q$-expansion of
$H_P^{(a)}\left({1\over  2};\tau\right)$
is always with integer powers of $q$ (upto an overall
factor $q^{-\frac{a^2}{4P}}$) while the sum of
$H_P^{(a)}\left({1+\tau\over 2}; \tau\right)$ and
$H_P^{(a)}\left({\tau\over 2} ; \tau \right)$ has either all integer
powers (for $a=\text{even}$) or half-integer powers
(for $a=\text{odd}$) up to an overall
factor
$q^{-\frac{a^2}{4P}+\frac{a}{2}}$.
Sum of $H_P^{(a)}$-functions have the $q$-expansion of the
form

\begin{equation}
  H_P^{(a)}\left({1\over 2};\tau\right)
  +H_P^{(a)}\left({1+\tau\over 2};\tau\right)
  +H_P^{(a)}\left({\tau\over 2};\tau\right)=
  q^{-\frac{a^2}{4(k+1)}} \,\left[ \alpha_{k+1}^{(a)}
    - \sum_{n \geq 1} A_{k+1,n}^{(a)} \, q^n\right].
\end{equation}
Substituting above into $Z^{(1)}_{X_k}$, we find
\begin{multline}
\label{degenerate-intermeadiate}
  c_k^{~-1}2^{-2k}
  Z^{(1)}_{X_k}(z;\tau)
  =
  3 \ch_{k,h=\frac{k}{4},\ell=0}^{\widetilde{R}}(z;\tau)
  \\
  -
  \sum_{a=1}^k
  q^{-\frac{a^2}{4(k+1)}} \,
  \left[
    \alpha_{k+1}^{(a)}
    - \sum_{n \geq 1} A_{k+1,n}^{(a)} \, q^n
  \right] \,
  \frac{
    \left[ \theta_{11}(z;\tau) \right]^2
  }{
    \left[ \eta(\tau) \right]^3
  } \,
  \chi_{k-1,\frac{a-1}{2}}(z;\tau) .
\end{multline}
Note that while terms with the coefficients $A_{k+1,n}^{(a)}$ give
massive representations 
with $h=n+k/4$, terms with  coefficients $\alpha_{k+1}^{(a)}$ are a
sum of massless representations  because
of the relation~\eqref{recursion_massless_R_tilde}.
%As in the case of level $k=1$  of~\eqref{l1_coefficient_massive}
%we conjecture that the expansion coefficients are all integers 
%\begin{equation}
 % A_{k+1,n}^{(a)} \in \mathbb{Z} .
%\end{equation}

We then rewrite (\ref{degenerate-intermeadiate}) using ~\eqref{recursion_massless_R_tilde} and find
\begin{equation}
  \label{decompose_X_k}
  \frac{1}{ c_k} \, Z^{(1)}_{X_k}(z;\tau)
  =
  \sum_{a=0}^{k}
  (-1)^a \,
  \gamma_{k,a} \,
  \ch_{k,\frac{k}{4},\frac{a}{2}}^{\widetilde{R}}(z;\tau)
  +
  \text{massive characters}.
\end{equation}
Here
a coefficient $\gamma_{k,a}$
of isospin-$a/2$  massless representation is given by
\begin{equation}
  \label{define_gamma_k}
  \gamma_{k,a}
  =
  2^{2k} \,
  \left(
    \alpha_{k+1}^{(a)}
    -2 \,
    \alpha_{k+1}^{(a+1)}
    +
    \alpha_{k+1}^{(a+2)} 
 \right) ,
\end{equation}
where  in our convention we have
\begin{equation*}
  \alpha_{k+1}^{(0)}=3 .
\end{equation*}
As we see from  Table~\ref{tab:gamma}  we have
$\gamma_{k,k}=1$, which counts
the multiplicity of massless representations with isospin $\frac{k}{2}$ in
the $R$-sector.
Since the spectral flow (in the left-moving sector) sends this representation to
identity representation in the NS sector while the right-moving sector
gives a degeneracy of $k+1$ for spin $k/2$ representation,  
we should multiply $Z^{(1)}_{X_k}$ by an overall factor $(k+1)$: then 
the multiplicity of identity representation in NS sector is adjusted to 1.
Thus the normalization parameter
in~\eqref{define_Z_X_k}  fixes to 
\begin{equation}
  \label{fix_c_k}
  c_k = k+1.
\end{equation}
We see  that \eqref{define_gamma_k} is computed for $a>0$  as
\begin{align}
  \gamma_{k,a}
  & =
  2^{2k}  \sum_{m=a}^k
  \frac{(-1)^{m+a}}{2^{2m-1}} \,
  \frac{a+1}{m+a+2}
  \begin{pmatrix}
    k \\
    m
  \end{pmatrix}
  \,
  \begin{pmatrix}
    2 \, m+1 \\
    m-a
  \end{pmatrix} 
  \nonumber
  \\
  & =
  \frac{2(a+1)}{k+a+2} \,
  \begin{pmatrix}
    2 \, k+1 \\
    k-a
  \end{pmatrix} ,
  \label{define_gamma_new}
\end{align}
and that
\begin{equation}
  \gamma_{k,0} =  C_{k+1}
  + 2^{2k+1},
\end{equation}
with the Catalan number
$C_n = \frac{1}{n+1} \,
\begin{pmatrix}
  2 n \\
  n
\end{pmatrix}$.
We then have
\begin{equation*}
  \frac{k+1}{a+1} \, \gamma_{k,a}  \in \mathbb{Z},
\end{equation*}
which gives the net number of massless representations 
with spin $\frac{k-a}{2}$ in the NS sector.
See Table~\ref{tab:gamma2}.

\begin{table}[htbp]
  \centering
  \begin{tabular}{c|rrrrrrrrrrr}
    $k \backslash a$ & 0 & 1 & 2 & 3 & 4 & 5 & 6 & 7 & 8 & 9 & 10
    \\
    \hline
    \hline
    1 & 10 & 1
    \\
    2 & 37 & 4 & 1
    \\
    3 & 142 & 14 & 6 & 1
    \\
    4 & 554 & 48 & 27 & 8 & 1 
    \\
    5 & 2180 & 165 & 110 & 44 & 10 & 1 
    \\
    6 & 8621 & 572 & 429 & 208 & 65 & 12 & 1
    \\
    7 & 34198 & 2002 & 1638 & 910 & 350 & 90 & 14 & 1
    \\
    8 & 135934 & 7072 & 6188 & 3808 & 1700 & 544 & 119 & 16 & 1
    \\
    9 & 541084 & 25194 & 23256 & 15504 & 7752 & 2907 & 798 & 152 &
    18 & 1
    \\
    10 & 2155938 & 90440 & 87210 & 62016 & 33915 & 14364 & 4655 & 1120
    & 189 & 20 & 1
    \\
    \hline
  \end{tabular}
  \caption{$ \gamma_{k,a}$ for $a=0,1,2,\dots,k$.}
  \label{tab:gamma}
\end{table}

\begin{table}[htbp]
  \centering
  \begin{tabular}{c|rrrrrrrrrrr}
    $k \backslash a$ & 0 & 1 & 2 & 3 & 4 & 5 & 6 & 7 & 8 & 9 & 10
    \\
    \hline
    \hline
    1 & 20 & 1
    \\
    2 & 111 & 6 & 1
    \\
    3 & 568 & 28 & 8 & 1
    \\
    4 & 2770 & 120 & 45 & 10 & 1 
    \\
    5 & 13080 & 495 & 220 & 66 & 12 & 1 
    \\
    6 & 60347 & 2002 & 1001 & 364 & 91 & 14 & 1
    \\
    7 & 273584 & 8008 & 4368 & 1820 & 560 & 120 & 16 & 1
    \\
    8 & 1223406 & 31824 & 18564 & 8568 & 3060 & 816 & 153 & 18 & 1
    \\
    9 & 5410840 & 125970 & 77520 & 38760 & 15504 & 4845 & 1140 & 190 &
    20 & 1
    \\
    10 & 23715318 & 497420 & 319770 & 170544 & 74613 & 26334 & 7315 & 1540
    & 231 & 22 & 1
    \\
    \hline
  \end{tabular}
  \caption{$\frac{k+1}{a+1} \, \gamma_{k,a}$ for
    $a=0,1,2,\dots,k$.}
  \label{tab:gamma2}
\end{table}

We note that
a series $\gamma_{k,a}$ for $a=1,2,\dots,k$ has a combinatorial
interpretation~\citep[A050156]{NSloanWWW};
we consider a set of  sequences of $0$'s and $1$'s,
\begin{equation}
  V(k)=
  \left\{
    v_1 v_2 \cdots v_{2k+1} 
    ~|~
    \#(\text{1's})=k+2,
    \#(\text{0's})=k-1
  \right\} .
\end{equation}
For each element $v_1 \cdots v_{2k+1} \in V(k)$,
we have a subsequence
$ v_1 v_2 \cdots v_h$ for $h=1,2,\dots,2\,k+1$.
Then we observe that
$\gamma_{k,a}$ denotes a cardinality of a subset $V(k)$ such that
\begin{equation*}
  \max_{h}
  \left\{
    \#(\text{1's})
    -
    \#(\text{0's})
  \right\}
  =a+2 .
\end{equation*}
From a viewpoint of statistical mechanics,
a set $V(k)$ of sequences is regarded as
a staircase walk which
starts from $(0,0)$ and ends at
$(k-1,k+2)$.
Here $0$ (resp. $1$) corresponds to 
a unit right-walk
$\rightarrow$
%$(1,0)$
(resp.   up-walk
$\uparrow$).
%$(0,1)$,
Then $\gamma_{k,a}$ coincides
with the number of staircase walks
which touch a line $y=x+a+2$ and do not go beyond the line.

%%%
\subsubsection{Other Cases}

Along the same strategy as the completely degenerate case, we can decompose  symmetric polynomials of the Jacobi theta
functions by using a suitable choice of points $w_1\cdots,w_{P-1}$.

Upon symmetrization, we obtain the following;
\begin{subequations}
  \label{general_elliptic_genus_0}
  \begin{enumerate}
    \renewcommand{\labelenumi}{(\alph{enumi})}
  \item $k_2 >k_3 > k_4$,
    \begin{multline}
      \label{general_genus_a}
      2^{2k_2-1}\left( \frac{\theta_{10}(z;\tau)}{\theta_{10}(0;\tau)} \right)^{2 k_2}
      \left( \frac{\theta_{00}(z;\tau)}{\theta_{00}(0;\tau)} \right)^{2 k_3}
      \left( \frac{\theta_{01}(z;\tau)}{\theta_{01}(0;\tau)} \right)^{2 k_4}
      +\text{other $5$ terms}
      \\
      =
      \sum_{a=0}^{k_2} (-1)^a \,
       \,
      \left(
        \gamma_{k_2,a}
        +
       2^{2k_2-2k_3}\gamma_{k_3,a}
        +
       2^{2k_2-2k_4}\gamma_{k_4,a}
      \right) \,
      \ch_{k,\frac{k}{4},\frac{a}{2}}^{\widetilde{R}}(z;\tau)
      +
      \text{massive characters},
    \end{multline}
    
  \item $k_2 = k_3 \neq k_4$,
    \begin{multline}
      \label{general_genus_b}
     2^{2k_2-1} \left(
        \frac{\theta_{10}(z;\tau)}{\theta_{10}(0;\tau)} \cdot
        \frac{\theta_{00}(z;\tau)}{\theta_{00}(0;\tau)}
      \right)^{2 k_2}
      \left(
        \frac{\theta_{01}(z;\tau)}{\theta_{01}(0;\tau)}
      \right)^{2 k_4}
      +\text{other $2$ terms}
      \\
      =
      \sum_{a=0}^{\max(k_2,k_4)} (-1)^a \,
      \left(
        \gamma_{k_2,a}
        +2^{2k_2-2k_4-1}
        \gamma_{k_4,a}
      \right) \,
      \ch_{k,\frac{k}{4},\frac{a}{2}}^{\widetilde{R}}(z;\tau)
      +
      \text{massive characters},
    \end{multline}

%   \item $k_2 > k_3 = k_4$,
%     \begin{multline}
%       \label{general_genus_c}
%       \left(
%         \frac{\theta_{10}(z;\tau)}{\theta_{10}(0;\tau)}
%         \cdot
%         \frac{\theta_{00}(z;\tau)}{\theta_{00}(0;\tau)}
%       \right)^{2 k_3}
%       \left(
%         \frac{\theta_{01}(z;\tau)}{\theta_{01}(0;\tau)}
%       \right)^{2 k_2}
%       +\text{other $2$ terms}
%       \\
%       =
%       \sum_{a=0}^{k_2} (-1)^a \,
%       \left(
%         \gamma_{k_2,a}
%         +
%         2 \, \gamma_{k_3,a}
%       \right) \,
%       \ch_{k,\frac{k}{4},\frac{a}{2}}^{\widetilde{R}}(z;\tau)
%       +
%       \text{massive characters},
%     \end{multline}

  \item $k_2=k_3=k_4$,
    \begin{multline}
      \label{general_genus_d}
     2^{2k_2} \left( \frac{\theta_{10}(z;\tau)}{\theta_{10}(0;\tau)}
        \cdot
        \frac{\theta_{00}(z;\tau)}{\theta_{00}(0;\tau)}
        \cdot
        \frac{\theta_{01}(z;\tau)}{\theta_{01}(0;\tau)}
      \right)^{2 k_2}
      \\
      =
      \sum_{a=0}^{k_2} (-1)^a \, \gamma_{k_2,a} \,
      \ch_{k,\frac{k}{4}, \frac{a}{2}}^{\widetilde{R}}(z;\tau)
      +
      \text{massive characters}.
    \end{multline}
  \end{enumerate}
\end{subequations}
Here $\gamma_{k,a}$ is defined by~\eqref{define_gamma_k} with
\begin{equation*}
  \alpha_{k+1}^{(0)}  = 1 ,
\end{equation*}
which coincides with~\eqref{define_gamma_new} for $a \geq 0$.
Normalization factors in the left hand sides
in~\eqref{general_elliptic_genus_0} are fixed so that
coefficients  of the massless character
$\ch_{k,\frac{k}{4},\frac{a}{2}}^{\widetilde{R}}(z;\tau)$ are integral.
% We stress that
% a fact 
% $\gamma_{k,a}=0$ for $a>k$
% proves that
% the isospin-$\frac{k}{2}$ massless character
% $\ch_{k,\frac{k}{4},\frac{k}{2}}^{\widetilde{R}}(z;\tau)$,
% which corresponds to
% the  character 
% $\ch_{k,\frac{k}{4},0}^{\widetilde{NS}}(z;\tau)$
% of identity representation in the NS sector,
% only appears in the degenerate case
% $(k_2, k_3, k_4)=(k,0,0)$
% %$k_2=k$, $k_3=k_4=0$
% studied before.
% Therefore we can  fix normalization factors only the degenerate
% case~\eqref{define_Z_X_k}
% as in~\eqref{fix_c_k}.
We emphasize that the massless isospin $k/2$ representation appears
only in the completely degenerate case $(k_2,k_3,k_4)=(k,0,0)$
for which the overall normalization can be fixed by requiring the multiplicity
to be one for the vacuum representation in NS sector.
We cannot fix
the overall normalization in the case of other configurations.  

%%%%%%%%%%%%%

\subsection{%\mathversion{bold}
  Example:
  Hyper-K{\"a}hler Manifold with $\dim_{\mathbb{C}}=4$}

We have two cases,
$(k_2,k_3,k_4)=(2,0,0)$ and
$(1,1,0)$.
The former case is~\eqref{define_Z_X_k} with $k=2$;
\begin{multline*}
  48 \,
  \left[
    \left(
      \frac{\theta_{10}(z;\tau)}{\theta_{10}(0;\tau)}
    \right)^{4}
    +
    \left(
      \frac{\theta_{00}(z;\tau)}{\theta_{00}(0;\tau)}
    \right)^{4}
    +
    \left(
      \frac{\theta_{01}(z;\tau)}{\theta_{01}(0;\tau)}
    \right)^{4}
  \right]
  \\
  =
  111 \,
  \ch_{2,\frac{2}{4},0}^{\widetilde{R}}(z;\tau)
  -
  12 \,
  \ch_{2,\frac{2}{4},\frac{1}{2}}^{\widetilde{R}}(z;\tau)
  +
  3 \,
  \ch_{2,\frac{2}{4},1}^{\widetilde{R}}(z;\tau)
  \\
  +
  \left(
    1872 \, q + 26070 \, q^2 + 213456 \, q^3 + 1311420 \, q^4+ \cdots
  \right) \, q^{-\frac{1}{12}} \, B_3^{(1)}(z;\tau)
  \\
  -
  \left(
    510 \, q + 12804 \, q^2 + 126360 \, q^3 + 841176 \, q^4+ \cdots
  \right) \, q^{-\frac{1}{3}} \, B_3^{(2)}(z;\tau) .
\end{multline*}
The latter case follows from~\eqref{general_genus_b} and reads as
\begin{multline*}
  2 \,
  \left[
    \left(
      \frac{\theta_{10}(z;\tau)}{\theta_{10}(0;\tau)} \cdot
      \frac{\theta_{00}(z;\tau)}{\theta_{00}(0;\tau)}
    \right)^{2}
    +
    \left(
      \frac{\theta_{00}(z;\tau)}{\theta_{00}(0;\tau)} \cdot
      \frac{\theta_{01}(z;\tau)}{\theta_{01}(0;\tau)}
    \right)^{2}
    +
    \left(
      \frac{\theta_{01}(z;\tau)}{\theta_{01}(0;\tau)} \cdot
      \frac{\theta_{10}(z;\tau)}{\theta_{10}(0;\tau)}
    \right)^{2}
  \right]
  \\
  =
  4 \,
  \ch_{2,\frac{2}{4},0}^{\widetilde{R}}(z;\tau)
  -
  \ch_{2,\frac{2}{4},\frac{1}{2}}^{\widetilde{R}}(z;\tau)
  +
  \left(
    16 \, q + 55 \, q^2 + 144 \, q^3 + 330 \, q^4 \cdots
  \right) \, q^{-\frac{1}{12}} \, B_3^{(1)}(z;\tau)
  \\
  +
  \left(
    10 \, q + 44 \, q^2 + 110 \, q^3 + 280 \, q^4 + \cdots
  \right) \, q^{-\frac{1}{3}} \, B_3^{(2)}(z;\tau) .
%  +
%  \text{massive characters}.
\end{multline*}
In the above decomposition the isospin $1/2$ massless representation
has a coefficient 1: we should multiply an overall factor 2 to account
for the degeneracy of the right-moving sector.
% We also should multiply
% by a factor 3 in order to account for the degeneracy for the isospin
% $1$ massive representations (terms proportional to $B^{(2)}_3)$.
Furthermore we require that all coefficients of massless and massive
characters should
be positive integers in the NS sector when we combine the above 2 cases.
Then 
the elliptic genus of hyper-K{\"a}hler manifold $X_2(n)$
with complex dimension 4  may be written as
\begin{multline}
  Z_{X_2(n)}(z;\tau)
  =
  48 \,
  \left[
    \left(
      \frac{\theta_{10}(z;\tau)}{\theta_{10}(0;\tau)}
    \right)^{4}
    +
    \left(
      \frac{\theta_{00}(z;\tau)}{\theta_{00}(0;\tau)}
    \right)^{4}
    +
    \left(
      \frac{\theta_{01}(z;\tau)}{\theta_{01}(0;\tau)}
    \right)^{4}
  \right]
  \\
  +
  4 \, n \,
  \left[
    \left(
      \frac{\theta_{10}(z;\tau)}{\theta_{10}(0;\tau)} \cdot
      \frac{\theta_{00}(z;\tau)}{\theta_{00}(0;\tau)}
    \right)^{2}
    +
    \left(
      \frac{\theta_{00}(z;\tau)}{\theta_{00}(0;\tau)} \cdot
      \frac{\theta_{01}(z;\tau)}{\theta_{01}(0;\tau)}
    \right)^{2}
    +
    \left(
      \frac{\theta_{01}(z;\tau)}{\theta_{01}(0;\tau)} \cdot
      \frac{\theta_{10}(z;\tau)}{\theta_{10}(0;\tau)}
    \right)^{2}
  \right] ,
\end{multline}
where $n$ is an  integer satisfying
\begin{equation*}
%  -2 \leq n \leq 8.
  -6 \leq n \leq 25.
\end{equation*}
We have
\begin{equation}
  \begin{aligned}
  \label{top-inv}
    Z_{X_2(n)}(z=0; \tau)
    & =
    12 \, (n + 12),
%    36 \, (n + 4),
    \\[2mm]
    Z_{X_2(n)}
    \left( z=\frac{1}{2};\tau \right)
    & =
%    12 \, (n+8)
    4 \, (n+24)
    +
    12288 \, q + 294912 \, q^2 + \cdots
    ,
    \\[2mm]
    Z_{X_2(n)}
    \left( z=\frac{1+\tau}{2};\tau \right)
    & =
    3 \, q^{-1} -
%    12 \, (n+2) \, q^{-\frac{1}{2}}
    4 \, (n+6) \, q^{-\frac{1}{2}}
    +
    828 + \cdots,
  \end{aligned}
  \end{equation}
  In the  mathematical literature, see for instance~\cite{Nieper-Wisskirchen},
two representative examples of complex 4-dimensional hyper-K\"ahler manifolds are discussed: 
Hilbert scheme of points on $K3$ surfaces $K^{[2]}$ and complex tori $A^{[[3]]}$.
Their topological invariants are given by
\begin{equation}
  \begin{array}{c|ccc}
    X& \chi_X & \sigma_X &\widehat{A}_X\\
    \hline \hline
    K^{[2]} & 324 & 156 & 3 \\
    A^{[[3]]} & 108 &  84 & 3
  \end{array}
\end{equation}
These values agree exactly with those of~\eqref{top-inv} for
%$n=5$ ($K^{[2]})$
$n=15$ ($K^{[2]})$
and
%$n=-1$ ($A^{[[3]]}$),
$n=-3$ ($A^{[[3]]}$),
respectively.
We in particular predict $\widehat{A}=3$ for any hyper-K\"ahler
manifold in 4-dimensions.

%%%%%%%%%%%%% 
\section{Concluding Remarks}

We have developed a method of improving the modular properties of BPS
representations in 
$\mathcal{N}=4$ SCA by taking an analogy with a theory of mock theta
functions. We have obtained a systematic treatment of character
decomposition which can be applied to study the elliptic genera of
hyper-K\"ahler manifolds. Our treatment of positivity and integrality
properties on the infinite series of massive representations is yet
incomplete. We would like to discuss these issues in detail and also
the case of higher dimensional hyper-K\"ahler manifolds  in 
forthcoming publications.

%%%%%%
\section*{Acknowledgments}
T.E. would like to thank Y.~Sugawara and A.~Taormina for discussions
and introduction to mathematical literature on mock theta functions.
K.H. would like to thank K.~Bringmann and J.~Lovejoy
for  useful communications.
This work is supported in part by Grant-in-Aid from the Ministry of
Education, Culture, Sports, Science and Technology of Japan.
%%%%%%%%%%%%%
%\newpage
\appendix
\section{Theta Functions}
\label{sec:theta}
\subsection{Jacobi Theta Function}
The Jacobi theta functions are defined by
\begin{align*}
%  \theta_1(z;\tau)
  \theta_{11}(z;\tau)
  & =
  \sum_{n \in \mathbb{Z}}
  q^{\frac{1}{2} \left( n+ \frac{1}{2} \right)^2} \,
  \E^{2 \pi \I \left(n+\frac{1}{2} \right) \,
    \left( z+\frac{1}{2} \right)
  }
  =
  - \I \, \theta_1(z; \tau) ,
%   \\
%   &  =
%   \I \, q^{\frac{1}{8}} \, \E^{\pi \I z} \,
%   \left(
%     q, \E^{-2 \pi \I z}, \E^{2 \pi \I z} \, q
%   \right)_\infty
  \\[2mm]
%
%  \theta_{2}(z;\tau)
  \theta_{10}(z;\tau)
  & =
  \sum_{n \in \mathbb{Z}}
  q^{\frac{1}{2} \left( n + \frac{1}{2} \right)^2} \,
  \E^{2 \pi \I \left( n+\frac{1}{2} \right) z}
  =
  \theta_2(z;\tau) ,
  \\[2mm]
%
%  \theta_3 (z;\tau)
  \theta_{00} (z;\tau)
  & =
  \sum_{n \in \mathbb{Z}}
  q^{\frac{1}{2} n^2} \,
  \E^{2 \pi \I  n  z}
  = \theta_3 (z;\tau) ,
  \\[2mm]
%
%  \theta_0 (z;\tau)
  \theta_{01} (z;\tau)
  & =
  \sum_{n \in \mathbb{Z}}
  q^{\frac{1}{2} n^2} \,
  \E^{2 \pi \I n \left( z+\frac{1}{2} \right) }
  =
  \theta_4(z;\tau) ,
\end{align*}
where we have also shown the relation to the conventional notations.
Under the $S$-transformation, we have
\begin{equation}
  \begin{pmatrix}
    \theta_{11}(z;\tau) \\
    \theta_{10}(z;\tau) \\
    \theta_{00}(z;\tau) \\
    \theta_{01}(z;\tau) 
  \end{pmatrix}
  =
  \sqrt{ \frac{\I}{\tau} } \, \E^{- \pi \I \frac{z^2}{\tau}} \,
  \begin{pmatrix}
    \I & & & \\
    & & & 1 \\
    & & 1 & \\
    & 1 & &
  \end{pmatrix} \,
  \begin{pmatrix}
    \theta_{11}\left( \frac{z}{\tau}; -\frac{1}{\tau} \right) \\
    \theta_{10}\left( \frac{z}{\tau}; -\frac{1}{\tau} \right) \\
    \theta_{00}\left( \frac{z}{\tau}; -\frac{1}{\tau} \right) \\
    \theta_{01}\left( \frac{z}{\tau}; -\frac{1}{\tau} \right) 
  \end{pmatrix} .
\end{equation}

We see that for $k \in \mathbb{Z}$
\begin{equation}
  \label{theta_power_expand}
  \begin{gathered}
    \left(
      \frac{\theta_{10}(z;\tau)}{\theta_{10}(0;\tau)}
    \right)^{2 k}
    =
    \left( \frac{\E^{\pi \I z}+ \E^{- \pi \I z}}{2} \right)^{2k}
    \,
    \left(
      1 
      +2 \, k \,
      \left(\E^{\pi \I z} - \E^{- \pi \I z} \right)^2 \, q
      + \cdots
    \right) ,
    \\[2mm]
    \left(
      \frac{\theta_{00}(z;\tau)}{\theta_{00}(0;\tau)}
    \right)^{2 k}
    =
    1 + 2 \, k \, \left( \E^{\pi \I z} - \E^{- \pi \I z} \right)^2 \,
    q^{\frac{1}{2}} + \cdots ,
    \\[2mm]
    \left(
      \frac{\theta_{01}(z;\tau)}{\theta_{01}(0;\tau)}
    \right)^{2 k}
    =
    1 - 2 \, k \, \left( \E^{\pi \I z} - \E^{- \pi \I z} \right)^2 \,
    q^{\frac{1}{2}} + \cdots .
  \end{gathered}
\end{equation}

%%%%
\subsection{Jacobi Form}

Jacobi form $\varphi(z;\tau)$ with weight-$k$ and index-$m$ has
following transformation formulae~\cite{EichZagi85};
\begin{equation}
  \begin{gathered}
    \varphi \left(
      \frac{z}{\tau} ; - \frac{1}{\tau}
    \right)
    =
    \tau^k \,
    \E^{ - 2 \pi \I m \frac{z^2}{\tau}} \,
    \varphi(z;\tau) ,
    \\[2mm]
    \varphi(z;\tau+1)
    =
    \varphi(z+1; \tau) = \varphi(z;\tau),
    \\[2mm]
    \varphi(z+\tau;\tau) 
    =
    q^{-m} \,
    \E^{-4 \pi \I m z} \,
    \varphi(z;\tau) .
  \end{gathered}
\end{equation}
% It is known~\cite{EichZagi85} that the space of Jacobi forms 
% with weight
% $k$ and index $m$  is isomorphic to a certain space of vector-valued
% modular forms of weight $k-\frac{1}{2}$.
%%%%
\subsection{Theta Functions and Characters}
We define the theta function for $a \mod 2 \, P$ by
(see, \emph{e.g.}, \cite{Shimura73a})
\begin{align}
  \vartheta_{P,a}(z;\tau)
  & =
  \vartheta
  \begin{bmatrix}
    \frac{a}{2 P} \\[1mm]
    0
  \end{bmatrix}
  \left( 2 \, P \, z ; 2 \, P \, \tau \right)
  \nonumber \\
  &  =
  \sum_{n \in \mathbb{Z}} q^{\frac{(2 P n +a)^2}{4 P}} \,
  \E^{2 \pi \I z (2 P n +a)} .
  \label{define_vartheta}
\end{align}
% By use of the Jacobi triple product identity, we  have an infinite
% product expression;
% \begin{align}
%   \vartheta_{P,a}(z;\tau)
%   & =
%   q^{\frac{a^2}{4 P}} \, \E^{2 \pi \I a z} \,
%   \left(
%     q^{2P} , -q^{P-a} \, \E^{-4 \pi \I P z},
%     -q^{P+a} \, \E^{4 \pi \I P z} ;
%     q^{2 P}
%   \right)_\infty
% \end{align}
A set of functions
$\left\{\vartheta_{P,a}(z;\tau)
  \
  |
  \
  a  \mod 2 \,P \right\}$
spans a ($2\, P$)-dimensional space
% $V_{2 P}$, which is the space
of
entire function $f(z)$ such that
\begin{align*}
  f(z+1) & = f(z) ,
  \\[2mm] 
  f(z+\tau) & = \E^{ - 4 \pi \I P z - 2 \pi \I P \tau} \, f(z) .
\end{align*}
%See \emph{e.g.} Ref.~\citenum{Hikam95e}.

The $S$-transformation formula is
\begin{align}
  \vartheta_{P,a}(z;\tau)
  & =
  \sqrt{\frac{\I}{\tau}} \,
  \frac{1}{\sqrt{2 \, P}} \, \E^{- \pi \I \frac{2 P}{\tau} z^2} \,
  \sum_{b=0}^{2 P-1}
  \E^{\frac{a b}{P} \pi \I} \,
  \vartheta_{P,b}\left(
    \frac{z}{\tau} ; - \frac{1}{\tau}
  \right)  .
%   \\
%   & =
%   \sqrt{\frac{\I}{\tau}} \,
%   \frac{1}{\sqrt{2 \, P}} \, \E^{- \pi \I \frac{2 P}{\tau} z^2} \,
%   \theta_{00} \left(
%     \frac{z}{\tau}+ \frac{a}{2 \, P} ; 
%     - \frac{1}{2 \, P \, \tau}
%   \right)
\end{align}
With a help of this identity, we have
\begin{equation}
  \left(
    \vartheta_{P,a}
    -
    \vartheta_{P,-a}
  \right) (z;\tau)
  = 
  \I \, \sqrt{\frac{\I}{\tau}} \, \E^{-2 \pi \I \frac{P}{\tau} z^2}
  \,
  \sum_{b=1}^{P-1}
  \mathbf{S}(P)_{ab} \,
  \left(
    \vartheta_{P,b}
    -
    \vartheta_{P,-b}
  \right) \left(
    \frac{z}{\tau};- \frac{1}{\tau}
  \right) ,
\end{equation}
where
\begin{equation}
  \mathbf{S}(P)_{ab} =
  \sqrt{\frac{2}{P}} \, \sin
  \left(\frac{a \, b}{P} \,  \pi \right) .
\end{equation}

The character of the level-$k$ SU(2) affine algebra is given
by~\cite{Kac90}
\begin{equation}
  \chi_{k,\ell}(z;\tau)
  =
  \frac{
    \vartheta_{k+2,2\ell+1} - \vartheta_{k+2,-2\ell-1}
  }{
    \vartheta_{2,1} - \vartheta_{2,-1}
  }(z;\tau) .
\end{equation}
Note that
the Macdonald--Weyl denominator identity proves
\begin{equation*}
  \left(
    \vartheta_{2,1} - \vartheta_{2,-1}
  \right) (z;\tau)
  =
  - \I \,
  \theta_{11}(2 \, z; \tau) .
%   q^{\frac{1}{8}} \, \E^{2 \pi \I z} \,
%   \left(
%     q, \E^{-4 \pi \I z} , q \, \E^{4 \pi \I z}
%   \right)_\infty
\end{equation*}
We define the weight-$3/2$ modular form from the SU(2) affine
character as
\begin{equation}
  \label{Psi_and_chi}
  \frac{
    \Psi_P^{(a)}(\tau)}{
    \left[ \eta(\tau) \right]^3
  }
  =
  \chi_{P-2, \frac{a-1}{2}}(0;\tau)
  =
  \frac{
    \vartheta_{P,a} - \vartheta_{P,-a}
  }{
    \vartheta_{2,1} - \vartheta_{2,-1}
  }(0;\tau) ,
\end{equation}
where $a\in\mathbb{Z}$ satisfying $0 < a < P$.
We see that
the modular form has a Fourier transformation as
\begin{equation}
  \Psi_P^{(a)}(\tau)
  =
  \frac{1}{2} \sum_{n \in \mathbb{Z}} n \,  \psi_{2P}^{(a)}(n) \, 
  q^{\frac{n^2}{4P}}  ,
  \label{Def_Psi}
\end{equation}
where we have used an odd periodic function
\begin{equation*}
  \psi_{2P}^{(a)}(n)
  =
  \begin{cases}
    \pm 1, & n=\pm a \mod 2P,
    \\[2mm]
    0, & \text{otherwise} .
  \end{cases}
\end{equation*}
See that, in case $P=2$,
we have
\begin{equation*}
  \Psi_2^{(1)}(\tau) =
  \left[ \eta(\tau) \right]^3 .
\end{equation*}
The modular transformation properties are summarized as
\begin{equation}
  \label{S_transform}
  \begin{gathered}
    \Psi_P^{(a)}(\tau)
    =
    \left(
      \frac{\I}{\tau}
    \right)^{3/2} \,
    \sum_{b=1}^{P-1} \mathbf{S}(P)_{ab} \,
    \Psi_P^{(b)}\left(-\frac{1}{\tau}\right) ,
    \\[2mm]
    \Psi_P^{(a)}(\tau+1)
    =
    \E^{\frac{a^2}{2P} \pi \I} \,
    \Psi_{P}^{(a)}(\tau) .
  \end{gathered}
\end{equation}

%%%%%%%%%%%%%%%

%%%
\section{Fourier Expansions of the Harmonic Maass Forms}
\label{sec:Fourier_Maass}

We give the Fourier expansion of the harmonic Maass
forms
$M_P^{(a)}(z;\tau)$~\eqref{define_M_P} at
$z\in \left\{ \frac{1}{2}, \frac{1+\tau}{2}, \frac{\tau}{2} \right\}$,
which are useful in the character decomposition of the elliptic
genera in terms of the level $k=P-1$ SCFT.
We have directly computed~\eqref{define_M_P} using \texttt{Mathematica}
and \texttt{Maple}.
\begin{scriptsize}
\begin{itemize}
\item $P=2$
  \begin{equation}
    \begin{aligned}
      \mu \left(\frac{1}{2};\tau \right)
      =
      h_2(\tau) \cdot \eta(\tau)
      & =
      \frac{q^{-\frac{1}{8}}}{4} \,
      \left[
        1 + 3 \, q - 7 \, q^2 + 14 \, q^3 - 21 \, q^4
        + \cdots
      \right]
      \\[2mm]
      \mu \left(\frac{1+\tau}{2};\tau \right)
      =
      h_3(\tau) \cdot \eta(\tau)
      & =
      q^{\frac{3}{8}} \,
      \left[
        2  - 6 \, q^{\frac{1}{2}} + 14 \, q - 28 \, q^{\frac{3}{2}}
        + 54 \, q^2 - 98 \, q^{\frac{5}{2}} + 168 \, q^3
        - \cdots
      \right]
      \\[2mm]
      \mu \left(\frac{\tau}{2};\tau \right)
      =
      h_4(\tau) \cdot \eta(\tau)
      & =
      q^{\frac{3}{8}} \,
      \left[
        -2  - 6 \, q^{\frac{1}{2}} - 14 \, q - 28 \, q^{\frac{3}{2}}
        - 54 \, q^2 - 98 \, q^{\frac{5}{2}} - 168 \, q^3
        - \cdots
      \right]
    \end{aligned}
  \end{equation}

\item $P=3$
  \begin{gather}
    \begin{aligned}
      H_3^{(1)}\left( \frac{1}{2}; \tau \right)
      & =
      \frac{q^{-\frac{1}{12}}}{8} \,
      \Biggl[
      3  + 8 q  -25 q^2 +
      72 q^3 -\cdots
      \Biggr]
      \\[2mm]
      H_3^{(1)}\left( \frac{1+\tau}{2}; \tau \right)
      & =
      q^{\frac{5}{12}} \,
      \Biggl[
      4 - 20 q^{\frac{1}{2}} + 80 q - 270 q^{\frac{3}{2}} + 812 q^2 -
      2228 q^{\frac{5}{2}} + 5680 q^3 - 13650 q^{\frac{7}{2}} + \cdots
      \Biggr]
      \\[2mm]
      H_3^{(1)}\left( \frac{\tau}{2}; \tau \right)
      & =
      q^{\frac{5}{12}} \,
      \Biggl[
      -4 - 20 q^{\frac{1}{2}} - 80 q - 270 q^{\frac{3}{2}} - 812 q^2 -
      2228 q^{\frac{5}{2}} - 5680 q^3 - 13650 q^{\frac{7}{2}} - \cdots
      \Biggr]
    \end{aligned}
    \\[2mm]
    \begin{aligned}
      H_3^{(2)}\left( \frac{1}{2}; \tau \right)
      & = \frac{q^{-\frac{1}{3}}}{16} \,
      \Biggl[
      1  + 10 q -  20 q^2
      + 40 q^3 +\cdots
      \Biggr]
      \\[2mm]
      H_3^{(2)}\left( \frac{1+\tau}{2}; \tau \right)
      & = q^{\frac{2}{3}} \,
      \Biggl[
      5 - 32 q^{\frac{1}{2}} + 134 q - 448 q^{\frac{3}{2}} + 1315 q^2
      - 3520 q^{\frac{5}{2}} + 8764 q^3 - 20608 q^{\frac{7}{2}} + \cdots
      \Biggr]
      \\[2mm]
      H_3^{(2)}\left( \frac{\tau}{2}; \tau \right)
      & = q^{\frac{2}{3}} \,
      \Biggl[
      5 + 32 q^{\frac{1}{2}} + 134 q + 448 q^{\frac{3}{2}} + 1315 q^2
      + 3520 q^{\frac{5}{2}} + 8764 q^3 + 20608 q^{\frac{7}{2}} + \cdots
      \Biggr]
    \end{aligned}
  \end{gather}

\item $P=4$
  \begin{gather}
    \begin{aligned}
      H_4^{(1)}\left( \frac{1}{2}; \tau \right)
      & =
      \frac{q^{-\frac{1}{16}}}{64} \Biggl[
      29  + 70 q - 258 q^2 +
      911 q^3 + \cdots
      \Biggr]
      \\[2mm]
      H_4^{(1)}\left( \frac{1+\tau}{2}; \tau \right)
      & =
      q^{\frac{7}{16}} \Biggl[
      6 - 42 q^{\frac{1}{2}} + 238 q - 1134 q^{\frac{3}{2}} + 4718 q^2 +
      59754 q^{\frac{5}{2}} - 188480 q^3 + \cdots
      \Biggr]
      \\[2mm]
      H_4^{(1)}\left( \frac{\tau}{2}; \tau \right)
      & =
      q^{\frac{7}{16}} \Biggl[
      -6 - 42 q^{\frac{1}{2}} - 238 q - 1134 q^{\frac{3}{2}} - 4718 q^2 -
      59754 q^{\frac{5}{2}} - 188480 q^3 - \cdots
      \Biggr]
    \end{aligned}
    \displaybreak[0]
    \\[2mm]
    \begin{aligned}
      H_4^{(2)}\left( \frac{1}{2}; \tau \right)
      & =
      \frac{q^{-\frac{1}{4}}}{8} \Biggl[
      1  + 8 q - 21 q^2 
      +56 q^3 + \cdots
      \Biggr]
      \\[2mm]
      H_4^{(2)}\left( \frac{1+\tau}{2}; \tau \right)
      & =
      q^{\frac{3}{4}} \Biggl[
      14 - 128 q^{\frac{1}{2}} + 762 q - 3584 q^{\frac{3}{2}} + 14434 q^2 +
      51840 q^{\frac{5}{2}} - 170212 q^3 + \cdots
      \Biggr]
      \\[2mm]
      H_4^{(2)}\left( \frac{\tau}{2}; \tau \right)
      & =
      q^{\frac{3}{4}} \Biggl[
      14 + 128 q^{\frac{1}{2}} + 762 q + 3584 q^{\frac{3}{2}} + 14434 q^2 +
      51840 q^{\frac{5}{2}} + 170212 q^3 + \cdots
      \Biggr]
    \end{aligned}
    \displaybreak[0]
    \\[2mm]
    \begin{aligned}
      H_4^{(3)}\left( \frac{1}{2}; \tau \right)    
      & =
      \frac{q^{-\frac{9}{16}}}{64} \Biggl[
      1  + 21 q - 7 q^2 
      - 77 q^3 + \cdots
      \Biggr]
      \\
      H_4^{(3)}\left( \frac{1+\tau}{2}; \tau \right)    
      & =
      q^{\frac{15}{16}} \Biggl[
      14 - 140 q^{\frac{1}{2}} + 852 q - 3990 q^{\frac{3}{2}} + 15836 q^2 +
      55890 q^{\frac{5}{2}} - 180298 q^3 + \cdots
      \Biggr]
      \\
      H_4^{(3)}\left( \frac{\tau}{2}; \tau \right)    
      & =
      q^{\frac{15}{16}} \Biggl[
      -14 - 140 q^{\frac{1}{2}} - 852 q - 3990 q^{\frac{3}{2}} - 15836 q^2 -
      55890 q^{\frac{5}{2}} - 180298 q^3 - \cdots
      \Biggr]
    \end{aligned}
  \end{gather}

\item $P=5$
  \begin{gather}
    \begin{aligned}
      H_5^{(1)}\left( \frac{1}{2}; \tau \right)    
      & =
      \frac{q^{-\frac{1}{20}}}{128} \Biggl[
      65 + 144 q - 591 q^2 + \cdots
      \Biggr]
      \\[2mm]
      H_5^{(1)}\left( \frac{1+\tau}{2}; \tau \right)    
      & =
      q^{\frac{9}{20}} \Biggl[
      8 - 72 q^{\frac{1}{2}} + 528 q - 3252 q^{\frac{3}{2}} + 16560 q^2
      - 71268 q^{\frac{5}{2}} + \cdots
      \Biggr]
      \\[2mm]
      H_5^{(1)}\left( \frac{\tau}{2}; \tau \right)    
      & =
      q^{\frac{9}{20}} \Biggl[
      -8 - 72 q^{\frac{1}{2}} - 528 q - 3252 q^{\frac{3}{2}} - 16560 q^2
      - 71268 q^{\frac{5}{2}} - \cdots
      \Biggr]
    \end{aligned}
    \displaybreak[0]
    \\[2mm]
    \begin{aligned}
      H_5^{(2)}\left( \frac{1}{2}; \tau \right)    
      & =
      \frac{q^{-\frac{1}{5}}}{256} \Biggl[
      46  + 315 q -980 q^2 +\cdots
      \Biggr]
      \\[2mm]
      H_5^{(2)}\left( \frac{1+\tau}{2}; \tau \right)    
      & =
      q^{\frac{4}{5}} \Biggl[
      27 - 320 q^{\frac{1}{2}} + 2468 q - 14976 q^{\frac{3}{2}} + 77022 q^2
      - 349248 q^{\frac{5}{2}} + \cdots
      \Biggr]
      \\[2mm]
      H_5^{(2)}\left( \frac{\tau}{2}; \tau \right)    
      & =
      q^{\frac{4}{5}} \Biggl[
      27 + 320 q^{\frac{1}{2}} + 2468 q + 14976 q^{\frac{3}{2}} + 77022 q^2
      + 349248 q^{\frac{5}{2}} + \cdots
      \Biggr]
    \end{aligned}
    \displaybreak[0]
    \\[2mm]
    \begin{aligned}
      H_5^{(3)}\left( \frac{1}{2}; \tau \right)    
      & =
      \frac{   q^{-\frac{9}{20}}}{128} \Biggl[
      5 + 80 q - 90 q^2 - \cdots
      \Biggr]
      \\[2mm]
      H_5^{(3)}\left( \frac{1+\tau}{2}; \tau \right)    
      & =
      q^{\frac{21}{20}} \Biggl[
      48 - 630 q^{\frac{1}{2}} + 4984 q - 30072 q^{\frac{3}{2}} + 151776 q^2
      - 669194 q^{\frac{5}{2}} + \cdots
      \Biggr]
      \\[2mm]
      H_5^{(3)}\left( \frac{\tau}{2}; \tau \right)    
      & =
      q^{\frac{21}{20}} \Biggl[
      -48 - 630 q^{\frac{1}{2}} - 4984 q - 30072 q^{\frac{3}{2}} - 151776 q^2
      - 669194 q^{\frac{5}{2}} - \cdots
      \Biggr]
    \end{aligned}
    \displaybreak[0]
    \\[2mm]
    \begin{aligned}
      H_5^{(4)}\left( \frac{1}{2}; \tau \right)    
      & =
      \frac{
        q^{-\frac{4}{5}}}{256} \Biggl[
      1 + 36 q  + 96 q^2 +\cdots
      \Biggr]
      \\[2mm]
      H_5^{(4)}\left( \frac{1+\tau}{2}; \tau \right)    
      & =
      q^{\frac{6}{5}} \Biggl[
      42 - 576 q^{\frac{1}{2}} + 4614 q - 27776 q^{\frac{3}{2}} + 138567 q^2
      - 597032 q^{\frac{5}{2}} + \cdots
      \Biggr]
      \\[2mm]
      H_5^{(4)}\left( \frac{\tau}{2}; \tau \right)    
      & =
      q^{\frac{6}{5}} \Biggl[
      42 + 576 q^{\frac{1}{2}} + 4614 q + 27776 q^{\frac{3}{2}} + 138567 q^2
      + 597032 q^{\frac{5}{2}} + \cdots
      \Biggr]
    \end{aligned}
  \end{gather}
\end{itemize}

\end{scriptsize}

%%%%

%%%%%%%%%%%%%% %ewpage
%\bibliographystyle{physics}
% %\bibliographystyle{amsalpha}
% %\bibliographystyle{JHEP}
% %\bibliographystyle{siam}
%\bibliographystyle{alphaKH}
% %ibliographystyle{klunum}
%%%%%%%%%%%%%%%%%%
%\bibliography{_def,gravity,square,math,ba,tba,math5,vm,square2,math4,qalg,math3,math2,poisson,geometry,soliton,cft,knot,tqft,comb,number}

\end{document}